# To be or not to be polar: the ferroelectric and antiferroelectric nematic phases

Ewan Cruickshank[a], Paulina Rybak[b], Magdalena Majewska[b], Shona Ramsay[a], Cheng Wang[c], Chenhui Zhu[c], Rebecca Walker[a], John M. D. Storey[a], Corrie T. Imrie[a], Ewa Gorecka[b], Damian Pociecha*[b]

[a] Dr. E. Cruickshank, S. Ramsay, Dr. R. Walker, Prof. J. M. D. Storey, Prof. C. T. Imrie
Department of Chemistry, School of Natural and Computing Sciences, University of Aberdeen, AB24 3UE, Scotland, United Kingdom.

[b] P. Rybak, Dr. M. Majewska, Prof. E. Gorecka, Prof. D Pociecha
University of Warsaw, Faculty of Chemistry, ul. Zwirki i Wigury 101, 02-089, Warsaw, Poland
Email: pociu@chem.uw.edu.pl

[c] Dr. C. Wang, Dr. C. Zhu
Advanced Light Source, Lawrence Berkeley National Laboratory, 1 Cyclotron Road, Berkeley, 94720 CA, USA.

**Abstract:** We report the properties of two new series of compounds that show the ferroelectric nematic phase in which the length of a terminal chain is varied. The longer the terminal chain, the weaker the dipole-dipole interactions of the molecules are along the director, and thus the lower the temperature at which the axially ferroelectric nematic phase is formed. For homologues of intermediate chain length, between the non-polar and ferroelectric nematic phases, there is a wide temperature range nematic phase with antiferroelectric character. The size of the antiparallel ferroelectric domains critically increases upon transition to the ferroelectric phase. In dielectric studies, both collective ('ferroelectric') and non-collective fluctuations are present, the 'ferroelectric' mode softens weakly at the N-$N_X$ phase transition because the polar order in this phase is weak. The transition to the $N_F$ phase is characterized by a much stronger lowering of the mode relaxation frequency and an increase in its strength, typical critical behavior is observed.

Ferroelectric materials have a spontaneous reversible electric polarization and show piezoelectric and pyroelectric properties ensuring their widespread use in leading-edge electronics such as actuators, sensors and memory elements.[1,2] In a liquid crystal, the switching of the electric polarization is coupled with the elastic or optical properties of the material, and this is highly desirable for applications in soft optoelectronic devices.[3] Liquid crystalline improper ferroelectric phases have been known for decades; ferroelectric smectic phases have been studied since the 1970s[4] and ferroelectric columnar phases since the 1990s[5]. It has been shown that due to the competing interactions within these phases, their structures are often complex, and only a relatively small number of ferroelectric, antiferroelectric and ferrielectric phases have been found[6]. Furthermore, these phases have found only very limited commercialization in, for example, liquid crystal on silicon (LCoS) displays. Their wider application potential has failed to be realized mainly due to the challenge of producing defect free large area samples.

Recently, a polar nematic phase was discovered[7,8] and later assigned as the ferroelectric nematic phase, $N_F$.[9] This is the least ordered polar liquid crystalline phase. In the conventional nematic phase, N, the rod-like molecules more or less align in a common direction known as the director described by a unit vector, **n**, whereas their centres of mass are distributed randomly such that the phase has a fluid character. The director possesses inversion symmetry, i.e. **n** = −**n**, and so the phase is non-polar. In the $N_F$ phase, however, the inversion symmetry is broken, *i.e.* **n** ≠ −**n**, and the phase becomes polar. It appears that this polar $N_F$ phase, in contrast to the previously studied smectic and columnar phases, is a proper ferroelectric phase, in which the polar order is induced due to dipole-dipole interactions and the polarization is found along the director.[9] The high fluidity of the $N_F$ phase combined with its polar properties immediately caught the attention of scientists around the world due not only to its huge application potential but also its fundamental significance as a spontaneously ferroelectric fluid. The $N_F$ phase became one of the hottest topics in liquid crystal research.[10–34] Owing to the fluid nature of the $N_F$ phase, a uniform polarization direction can be obtained in large areas - a key to realizing its application potential.[35] However, the question arises whether the competitive interactions that drive the formation of the $N_F$ phase can also lead to other complex structures as is the case with the improper ferroelectric liquid crystal phases.

To date there have been some 200 mesogens reported which exhibit the $N_F$ phase but in general these materials are designed using three archetypal architectures which appear to have somewhat similar properties despite having differing chemical structures (Fig. 1).[8,11,36] These materials all have a large longitudinal dipole moment giving strong dipole-dipole interactions and also possess some degree of lateral bulk thought to inhibit anti-parallel correlations between molecules.[37,38]

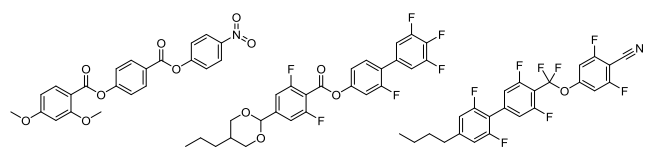

**Figure 1.** Molecular structures of archetypal ferroelectric nematogens: (left) RM734, (middle) DIO and (right) UUQU-4-N.

We report here two homologous series based on RM734[36] (Fig. 2), which both have strong longitudinal dipole moments (~11-12D), but in which the lateral methoxy group has been moved from the terminal to the central phenyl ring and a hydrogen *ortho* to the terminal nitro group is replaced by a fluorine atom. The series differ in the nature of the terminal chain; the *n*OEC3F series contains an alkyloxy chain and the *n*EC6F series an alkyl chain. For both series we report the change in behavior on extending the length of the terminal chain. A detailed description of the preparation of both these series, including the structural characterization data for all intermediates and final products, is provided in the Supplementary Information.

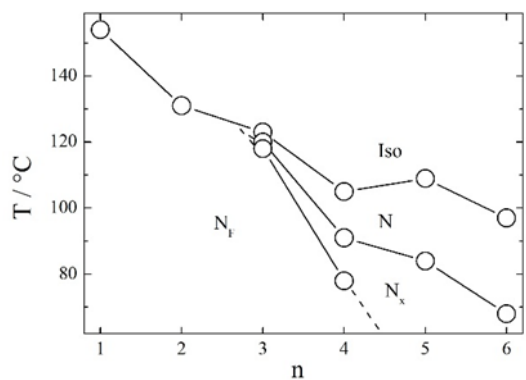

**Figure 3.** Phase diagram for the *n*EC6F series of ferroelectric nematogens with molecular structure shown in Fig.2. The conventional non-polar nematic phase is represented by N, the antiferroelectric nematic phase by $N_X$ and the polar ferroelectric nematic phase by $N_F$.

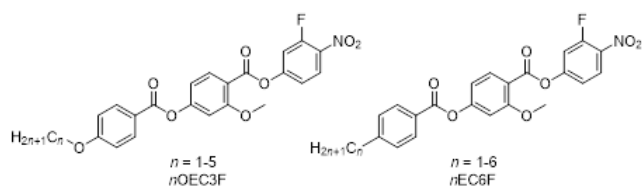

**Figure 2.** The molecular structures of: (left) the *n*OEC3F and (right) the *n*EC6F series where *n* refers to the number of carbon atoms in the terminal chain.

## Results and Discussion

We have shown previously that increasing the length of a lateral alkyloxy chain for RM734-type materials destabilizes the ferroelectric properties.[13,23,24,26] However the N-$N_F$ phase transition temperature decreases less than the Iso-N transition temperature, such that for most homologues in which there is a fluorine atom *ortho* to the terminal nitro group, a direct Iso-ferroelectric nematic phase transition is observed. In contrast, here we observed that increasing the length of the terminal alkyl chain only weakly affects the clearing temperatures, whereas the stability of ferroelectric $N_F$ phase strongly decreases in favor of the N and intermediate $N_X$ phases (Fig. 3). For the *n*EC6F series, when *n* = 1 and 2 a direct Iso-$N_F$ transition is seen, for *n* = 3 and 4 the sequence Iso-N-$N_X$-$N_F$ is observed, and for homologues *n* = 5 and 6 an Iso-N-$N_X$ phase sequence is found. The longest homologues crystallized close to room temperature, but without first entering the $N_F$ phase (Fig. 3). Such phase behavior is expected given that increasing the length of the terminal alkyl chain decreases the dipole-dipole interactions along the director, and thus the tendency to form an axially ferroelectric arrangement of dipole moments diminishes. The stabilization of the $N_X$ phase over a broad temperature range for the *n* = 5 and 6 homologues, offers the possibility for a detailed characterization of this phase. This is particularly important since the structure of $N_X$ phase is still under debate. Chen *et al* suggested that the phase has only short-range order regarding molecular positions, but shows a regular array of antiferroelectric domains along the direction perpendicular to the director.[39] To date, the structure of the $N_X$ phase (referred to by Chen *et al* as $SmZ_A$) was confirmed by X-ray diffraction studies in only a single compound, namely DIO (Fig. 1).

For the studied materials, the N-$N_X$ phase transition is weakly first order. It is accompanied by only a small, step-like increase of optical birefringence, of less than 0.001 for 4EC6F (Fig. 4a), and this decreases on increasing the terminal alkyl chain length. Thus, one can assume that the orientational order of the molecules remains similar in the N and $N_x$ phases.

In optical studies performed using planar aligned cells, there is clearly a transition detected at the temperature described by the birefringence measurements. In the N phase there is a uniform texture observed, and on entering the $N_X$ phase, the flickering characteristic to non-polar phases ceases and chevron-like defects appear a few degrees below phase transition (Fig. 4). The dielectric studies performed for homologue 4EC6F shows a weak dielectric mode in the N phase (with a relaxation frequency ~$10^5$ Hz) that continuously slows down but increases in strength through the entire range of the N and $N_X$ phases. The N-$N_X$ phase transition is marked by only a slight change in the value of the mode strength (Fig. 5a). This mode might be ascribed to the non-collective rotations of molecules with strong dipole moments around their highest inertia axis. Entering the $N_F$ phase, there is a dramatic change in the dielectric response and a very strong, low relaxation frequency (~$10^3$Hz) mode appears. Furthermore, in optical studies there is also a clear transition observed upon entry to the $N_F$ phase with the emergence of a blocky type texture with some focal conic-like defects (Fig. 4).[28] In the $N_X$ phase under a weak bias electric field, the relaxation mode due to non-collective molecular rotations is quenched and the mode due to the collective ferroelectric fluctuation is excited. Suppressing the non-collective fluctuations by a bias electric field allows us to follow the evolution of the 'ferroelectric' mode (~$10^3$ Hz) over the whole temperature range (Fig. 5b,c). This mode slightly 'softens' at the N-$N_X$ phase transition as the polar order in this phase is weak, but the transition to the $N_F$ phase is marked by a much stronger critical lowering of the mode frequency and an increase of its strength. Therefore, typical critical behavior is observed.

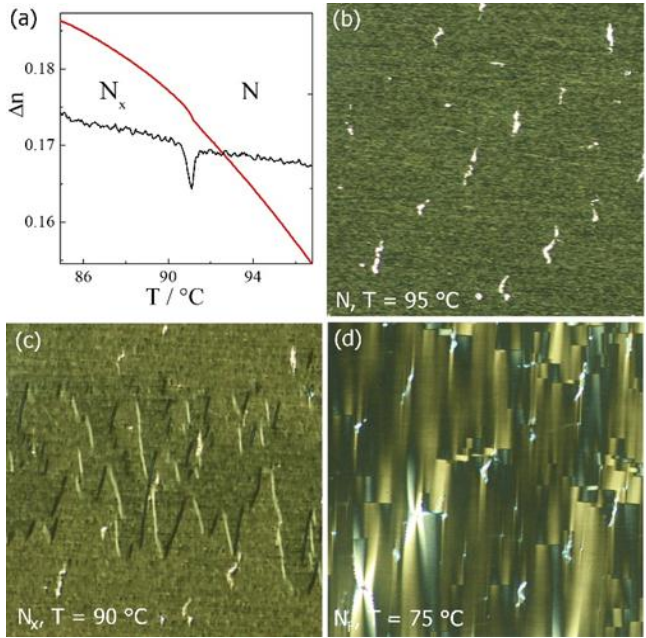

synchrotron source, in addition to the diffuse signal due to the short-range positional order of the molecules, which is typical for the nematic phase, there was also a separate sharp, machine resolution limited signal (Fig. 6). This observation proved that there was long range ordering within the phase due to the periodic structure of the antiferroelectric domains. The low intensity of the signal shows that the related electron density modulation is very weak. The signal position depends on the terminal alkyl chain such that higher values of $n$ give shorter periodicities, being, deep within the $N_X$ phase, around 75 Å, 50 Å and 40 Å (just 20-10 molecular widths) for 4EC6F, 5EC6F and 6EC6F respectively. It would appear that the size of the domains defining the $N_X$ structure can be correlated with the tendency to form the $N_F$ phase. The size of the domains increases on approaching the transition to the $N_F$ phase, and this tendency is clearly seen for the 4EC6F. However, for 5EC6F a much weaker and non-monotonic temperature dependence is observed. For 6EC6F, which is the homologue with the longest terminal alkyl chain length, the size of the domains monotonically decreases on cooling, suggesting that for this compound the tendency to form the ferroelectric nematic phase is very weak, and therefore does not influence the domain size.

**Figure 4.** (a) Optical birefringence (red line) of 4EC6F measured with red light ($\lambda$ = 690 nm) across the N-$N_X$ phase transition. The black line shows the derivative d($\Delta$n)/dT, and this clearly shows the transition temperature associated with the $N_X$ phase. (b-d) Optical textures of N, $N_X$ and $N_F$ phases observed between crossed polarizers in a 1.8-μm-thick cell with planar anchoring, with the chevron defects and focal conic-like defects appearing in the $N_X$ and $N_F$ phases, respectively.

For 5EC6F resonant soft x-ray scattering studies (RSoXS) were also conducted. The diffraction signal at the resonance condition was sensitive to the orientation of the molecules unlike conventional x-ray diffraction. The RSoXS signal was found at a periodicity double that detected in the SAXS measurements (Fig. 6), and this clearly confirms that structure is related to an antiparallel orientation of molecules in neighboring domains.

In order to probe the structure of the $N_X$ phase, small angle x-ray diffraction studies (SAXS) were performed. Using a strong

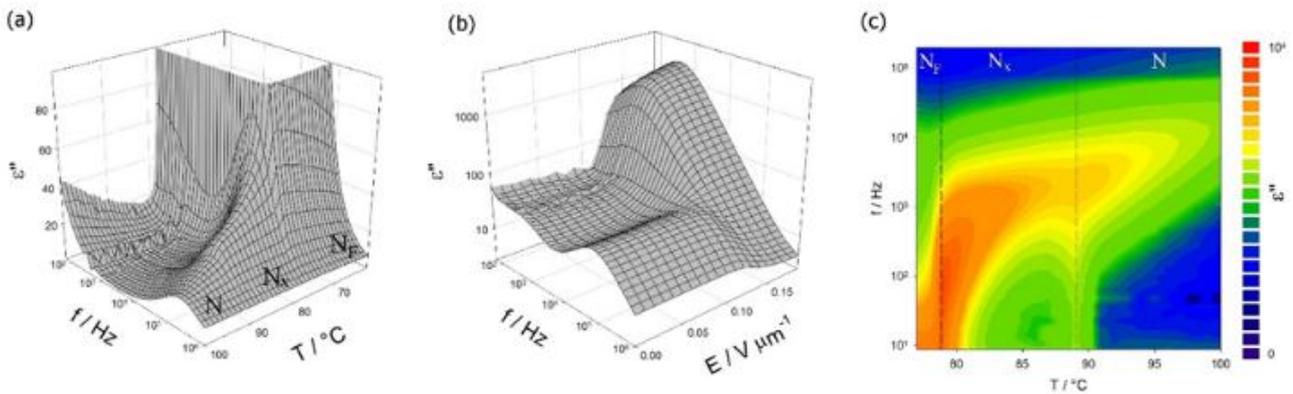

**Figure 5.** The imaginary part of dielectric susceptibility measured for 4EC6F ($n$ = 4): (a) the temperature and frequency dependence across the N-$N_X$-$N_F$ phase sequence, (b) the frequency and bias field dependence in the $N_X$ phase (at T = 80 °C); where the high-frequency mode is suppressed and the lower-frequency ferroelectric mode is excited above threshold field of 0.12 V / μm; and (c) a map showing the evolution of the 'ferroelectric' mode vs. temperature and frequency. The measurements were performed under a bias electric field of 0.3 V / μm.

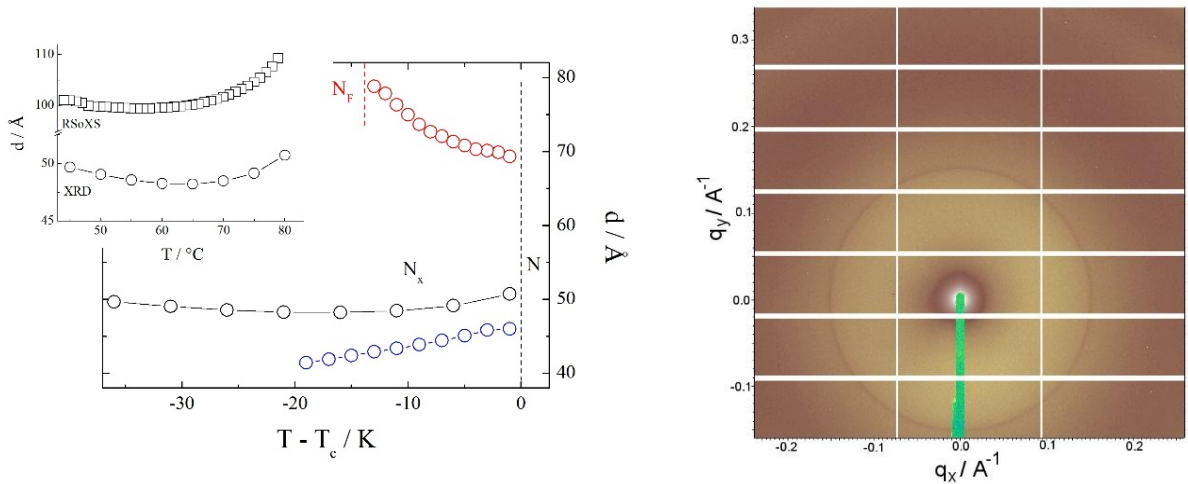

**Figure 6.** (left) Periodicity of the antiferroelectric domain structure in the NX phase (d) vs. temperature for the *n*EC6F series with *n* = 4 (red), *n* = 5 (black) and *n* = 6 (blue) measured using X-ray diffraction. In the inset: comparison of the periodicities deduced from non-resonant (XRD) and resonant (RSoXS) studies; (right) the 2D XRD pattern registered for *n* = 6 in the $N_X$ phase, at T = 40 °C. The sharp signal at q = 0.15 Å$^{-1}$ is due to periodic structure of antiferroelectric domains, while the diffused signal centered at q = 0.26 Å$^{-1}$ reflects the short-range positional order of molecules along the director of the nematic phase.

For comparison we studied the homologous series *n*OEC3F, in which the terminal alkyl chain is replaced by an alkyloxy chain. Although in general the stability of the liquid crystalline phases is increased by introducing an oxygen atom between a terminal alkyl chain and the mesogenic unit, the tendency to form the polar $N_F$ phase was diminished (Fig. S1). This may be somewhat surprising considering the average overall molecular dipole of the *n*OEC3F series is 13.0 D compared to 12.0 D for the *n*EC6F series, (Fig. 7). Apparently, larger longitudinal dipole moments are not exclusively the driving force for the formation of the $N_F$ phase. Within the framework of the model of the $N_F$ phase proposed by Madhusuhana, the molecules are described by longitudinal surface charge density waves which interact to prevent the formation of antiparallel structures.[40] In order to stabilize the ferroelectric nematic phase and promote the parallel alignment of the calamitic molecules, the amplitude of the charge density waves at either end of the molecule should be reduced. We have reported previously that this may be achieved using a fluorine atom at the *ortho* position to the terminal nitro group rather than a hydrogen atom, in order to reduce the electron density associated with the nitro group.[23,24] In this case, however, we are instead changing the electron density associated with the ring to which the terminal alkyl or alkyloxy chain is attached. An alkyloxy chain is a stronger activating functional group when compared to an alkyl chain due to its enhanced electron donating character and this means that there is a greater electron density in the terminal ring of the *n*OEC3F series compared to the *n*EC6F series (Fig. 7).[41] This increase in electron density will cause the amplitude of the surface charge density wave to increase for the *n*OEC3F series and, hence the temperatures of the N-$N_F$ and N-$N_X$ phase transitions show a more pronounced decrease on the elongation of terminal chain for the alkyloxy derivatives when compared to their alkyl counterparts. This decrease may also be attributed, to some extent, to shape effects, given that the alkyloxy chain lies more or less in the plane of the mesogenic unit to which it is attached, whereas the alkyl chain protrudes at an angle, and this will also disrupt the anti-parallel correlations between the molecules (Fig. 7). This weaker tendency to form ferroelectric phases in the *n*OEC3F series was also confirmed by X-ray diffraction studies performed for the material 3OEC3F. The periodicity of the antiferroelectric domain structure in the $N_X$ phase varied from 45 to 42 Å (65 - 45 °C), which is less than that found for the alkyl terminated analogue 4EC6F, having the same total length of terminal chain.

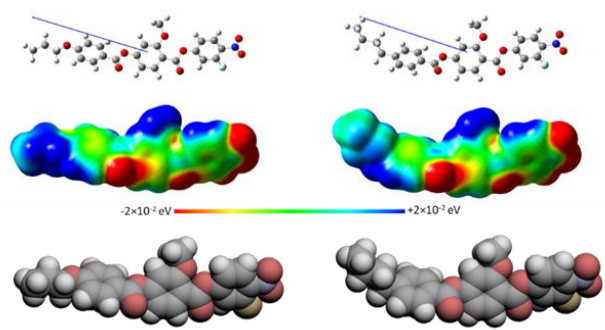

**Figure 7.** Molecular modelling of: (left) 3OEC3F and (right) 4EC6F calculated at the B3LYP/6-31(d) level of theory. The molecules are visualized using: (top) ball and stick models, (middle) electrostatic potential surfaces and (bottom) space-filling models. The arrow indicates the direction of the calculated dipole moment, with the head representing positive charge moving to the base which is negative.

The dielectric measurements for 3OEC3F revealed two clear relaxation modes in the whole temperature range of the $N_X$ and N phases (Fig. S2). As described for 4EC6F, upon the application of a bias electric field in the $N_X$ phase, the non-collective, higher frequency mode is quenched and instead the 'ferroelectric mode' is excited. In 3OEC3F, the temperature evolution of this 'ferroelectric mode' under a bias field showed a very weak softening behavior at the N-$N_X$ phase transition and this softening was much less pronounced than observed for the analogous compound 4EC6F. Such behavior is consistent with the optical observations, as there was a nearly smooth evolution of optical birefringence across the N-$N_x$ phase transition (Fig. S3), revealing its nearly continuous character and the weak polar ordering in the domains.

## Conclusion

In conclusion, the results obtained indicate that the $N_x$ phase is built from small polar regions, which form a regular antiferroelectric structure. The dielectric response measured indicates that the polar order in these regions is weak, and that the phase transition to the conventional non-polar nematic is very weakly first order. The periodicity of the antiferroelectric domains array in the $N_x$ phase increases with increasing ferroelectric interactions in the system, and the closer it is to the transition to the ferroelectric nematic phase, the wider these domains become. The question remains what causes the density modulation responsible for the weak X-ray signal in the $N_x$ phase? It is possible that either the domain walls have slightly different densities than the ferroelectric domains, or that the polarization splays or its magnitude is modulated across the domain, leading to slightly different densities at the domain boundaries.

## Acknowledgements

The research was supported by the National Science Centre (Poland) under the grant no. 2021/43/B/ST5/00240. C.T.I. and J.M.D.S. acknowledge the financial support of the Engineering and Physical Sciences Research Council under the grant no. EP/V048775/1. The beamlines 7.3.3 and 11.0.1.2 at the Advanced Light Source at the Lawrence Berkeley National Laboratory is supported by the Director of the Office of Science, Office of Basic Energy Sciences, of the U.S. Department of Energy under Contract No. DE-AC02- 05CH11231.

**Keywords:** Ferroelectric Nematic Phase, Antiferroelectric Nematic Phase, RSoXS, Dielectric Constant, Liquid Crystals

# Experimental

### Reagents

All reagents and solvents that were available commercially were purchased from Sigma Aldrich, Fisher Scientific or Fluorochem and were used without further purification unless otherwise stated.

### Thin Layer Chromatography

Reactions were monitored using thin layer chromatography, and the appropriate solvent system, using aluminium-backed plates with a coating of Merck Kieselgel 60 F254 silica which were purchased from Merck KGaA. The spots on the plate were visualised by UV light (254 nm).

### Column Chromatography

For normal phase column chromatography, the separations were carried out using silica gel grade 60 Å, 40-63 µm particle size, purchased from Fluorochem and using an appropriate solvent system.

### Structure Characterisation

All final products and intermediates that were synthesised were characterised using $^1$H NMR, $^{19}$F NMR, $^{13}$C NMR and infrared spectroscopies. The NMR spectra were recorded on a 400 MHz Bruker Avance III HD NMR spectrometer. The infrared spectra were recorded on a Perkin Elmer Spectrum Two FTIR with an ATR diamond cell.

### Purity Analysis

In order to determine the purity of the final products, high-resolution mass spectrometry was carried out using a Waters XEVO G2 Q-Tof mass spectrometer by Dr. Morag Douglas at the University of Aberdeen.

### Birefringence

The optical retardation was measured with a setup consisting of a photoelastic modulator (PEM-90, Hinds), halogen lamp (Hamamatsu LC8) equipped with a set of narrow bandpass filters as a light source, and a photodiode (FLC Electronics PIN-20). The measured intensity of the transmitted light was de-convoluted with a lock-in amplifier (EG&G 7265) into 1f and 2f components to yield a retardation induced by the sample. Glass cells with thickness 1.6 µm and surfactant assuring planar anchoring condition were used.

### Dielectric Measurments

The complex dielectric permittivity was measured in 1 Hz – 10 MHz frequency (f) range using Solatron 1260 impedance analyzer. Material was placed in glass cells with ITO or Au electrodes (and no polymer alignment layer to avoid influence of high capacitance of thin polymer layer) and thickness ranging from 5 to 10 microns. The relaxation frequency, $f_r$, and dielectric strength of the mode Δε, were evaluated by fitting the complex dielectric permittivity to Cole-Cole formula.

**X-ray diffraction studies**

Were performed at the Advanced Light Source, Lawrence Berkeley National Laboratory. Diffraction at small angle range were carried out on the SAXS beam line (7.3.3) at the energy of incident beam 10 keV. Samples were prepared in thin-walled glass capillaries or placed on heating plate as droplets. The scattering intensity was recorded using the Pilatus 2M detector, placed at the distance 2575 mm from the sample. The resonant x-ray scattering was performed on the soft x-ray beam line (11.0.1.2). The energy of incident beam was tuned to the K-edge of carbon absorption (283 eV). Samples with thickness lower than 1 µm were placed on a TEM grid. The scattering intensity was recorded using the Princeton PI-MTE CCD detector.

**Molecular modelling**

The geometric parameters of the *n*OEC3F and *n*EC6F series were calculated with quantum mechanical DFT calculations using Gaussian09 software.[1] Optimisation of the molecular structures was carried out at the B3LYP/6-31G(d) level of theory. A frequency check was used to confirm that the minimum energy conformation found was an energetic minimum. Visualisations of electronic surfaces and ball-and-stick models were generated from the optimised geometries using the GaussView 5 software, specifically the electronic surfaces were calculated using the cubegen utility in GaussView. Visualisations of the space-filling models were produced post-optimisation using the QuteMol package.[2]

**Polarised Optical Microscopy**

Optical studies were performed by using a Zeiss Axio Imager A2m polarising light microscope, equipped with a Linkam heating stage or using a Olympus BH2 polarising light microscope equipped with a Linkham TMS 92 hot stage. Samples were prepared in commercial cells (AWAT) of various thickness (1.5 – 20 µm) with ITO electrodes and planar alignment or in commercial cells purchased from INSTEC with a cell thickness of 2.9 – 3.5 µm and also planar alignment. The optical microscopic image was analyzed (director field and birefringence) with ABRIO system.

**Differential scanning calorimetry**

The phase behaviour of the materials was studied by differential scanning calorimetry performed using Mettler Toledo DSC1 or DSC3 differential scanning calorimeters equipped with TSO 801RO sample robots and calibrated using indium and zinc standards. Heating and cooling rates were 10 °C min$^{-1}$, with a 3-min isotherm between either heating or cooling, and all samples were measured under a nitrogen atmosphere. The enatiotropic transition temperatures and associated enthalpy changes were extracted from the heating traces wheress the monotropic transition temperatures and associated enthalpy changes were extracted from the cooling traces.

## Synthesis and Analytical Data

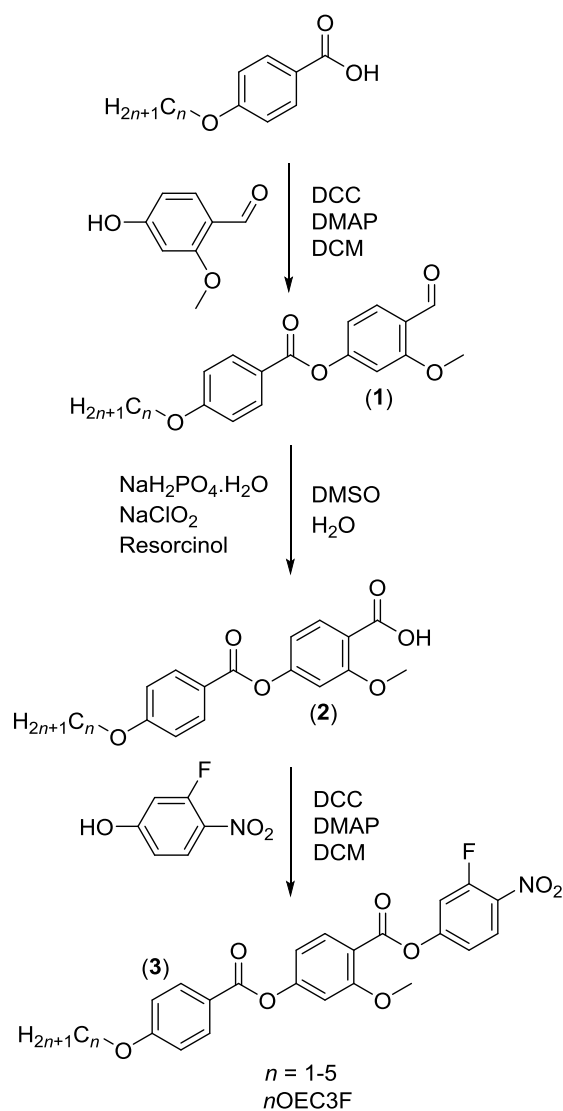

### *4-Formyl-3-methoxyphenyl 4-alkyloxybenzoates (1)*

To a pre-dried flask flushed with argon, 4-alkyloxybenzoic acid of the appropriate chain length (1 eq), 4-hydroxy-2-methoxybenzaldehyde (1.1 eq) and 4-dimethylaminopyridine (0.13 eq) were added. The solids were solubilised with dichloromethane (100 mL) and tetrahydrofuran (20 mL) while being stirred for 10 min before *N,N'*-dicyclohexylcarbodiimide (1.3 eq) was added to the flask and the reaction was allowed to proceed overnight. The quantities of the reagents used in each reaction are listed in **Table S1**. The extent of the reaction was monitored by TLC using an appropriate solvent system (RF values quoted in the product data). The precipitate which formed was removed by vacuum filtration and the filtrate collected. The collected solvent was evaporated under vacuum to leave a solid which was recrystallised from hot ethanol (200 mL).

**Table S1.** Quantities of reagents used in the syntheses of the 4-formyl-3-methoxyphenyl 4-alkyloxybenzoates (**1**)

| $n$ | 4-Alkyloxybenzoic Acid | 4-Hydroxy-2-methoxybenzaldehyde | 4-Dimethylaminopyridine | $N,N'$-Dicyclohexylcarbodiimide |
|---|---|---|---|---|
| 1 | 3.00 g, 0.0197 mol | 3.30 g, 0.0217 mol | 0.313 g, 2.56×10$^{-3}$ mol | 5.28 g, 0.0256 mol |
| 2 | 3.00 g, 0.0181 mol | 3.03 g, 0.0199 mol | 0.287 g, 2.35×10$^{-3}$ mol | 4.85 g, 0.0235 mol |
| 3 | 3.00 g, 0.0166 mol | 1.54 g, 0.0183 mol | 0.264 g, 2.16×10$^{-3}$ mol | 4.46 g, 0.0216 mol |
| 4 | 3.00 g, 0.0150 mol | 2.51 g, 0.0165 mol | 0.238 g, 1.95×10$^{-3}$ mol | 4.02 g, 0.0195 mol |
| 5 | 3.00 g, 0.0144 mol | 2.40 g, 0.0158 mol | 0.228 g, 1.87×10$^{-3}$ mol | 3.86 g, 0.0187 mol |

*1.1 4-Formyl-3-methoxyphenyl 4-methoxybenzoate*

White solid. Yield: 4.54 g, 80.5 %. RF: 0.542 (40 % ethyl acetate:60 % 40:60 petroleum ether). M.P = 145 °C

$v_{max}$/cm$^{-1}$: 2844, 1735, 1678, 1602, 1589, 1581, 1511, 1496, 1475, 1456, 1416, 1401, 1315, 1249, 1194, 1171, 1154, 1118, 1101, 1054, 1024, 1006, 946, 878, 841, 821, 807, 785, 757, 734, 686, 666, 639, 624, 605, 562, 509, 501, 463, 436, 405

$\delta_H$/ppm (400 MHz, CDCl$_3$): 10.40 (1 H, s, (C=O)-H), 8.12 (2 H, d, J 8.9 Hz, Ar-H), 7.87 (1 H, d, J 8.5 Hz, Ar-H), 6.97 (2 H, d, J 8.9 Hz, Ar-H), 6.88 (2 H, m, Ar-H), 3.91 (3 H, s, O-CH$_3$), 3.88 (3 H, s, O-CH$_3$)

$\delta_C$/ppm (100 MHz, CDCl$_3$): 188.63, 164.24, 164.13, 162.84, 157.24, 132.42, 129.82, 122.55, 121.10, 114.36, 114.00, 105.82, 55.92, 55.57

*1.2 4-Formyl-3-methoxyphenyl 4-ethoxybenzoate*

White solid. Yield: 3.58 g, 65.9 %. RF: 0.781 (60 % ethyl acetate:40 % 40:60 petroleum ether). M.P = 134 °C

$v_{max}$/cm$^{-1}$: 2979, 2897, 1728, 1680, 1604, 1589, 1514, 1489, 1466, 1450, 1423, 1414, 1387, 1254, 1190, 1121, 1097, 1068, 1037, 1025, 946, 921, 877, 853, 836, 807, 789, 758, 739, 690, 673, 626, 614, 563, 511, 500, 474, 461, 447, 423, 412, 405

$\delta_H$/ppm (400 MHz, DMSO-d$_6$): 10.31 (1 H, s, (C=O)-H), 8.08 (1 H, d, J 8.7 Hz, Ar-H), 7.78 (1 H, d, J 8.4 Hz, Ar-H), 7.21 (1 H, d, J 1.9 Hz, Ar-H), 7.12 (1 H, d, J 8.7 Hz, Ar-H), 7.00 (1 H, dd, J 8.4 Hz, 1.9 Hz, Ar-H), 7.00 (1 H, d, J 2.3 Hz, Ar-H), 4.15 (2 H, quart, 7.0 Hz, O-CH$_2$-CH$_3$), 3.91 (3 H, s, O-CH$_3$), 1.36 (3 H, t, 7.0 Hz, O-CH$_2$-CH$_3$)

$\delta_C$/ppm (100 MHz, DMSO-d$_6$): 188.35, 163.75, 163.37, 162.72, 157.16, 132.33, 129.33, 122.12, 120.42, 114.83, 114.80, 107.13, 63.88, 56.52, 14.56

### *1.3 4-Formyl-3-methoxyphenyl 4-propoxybenzoate*

Off-white solid. Yield: 2.17 g, 70.6 %. RF: 0.611 (40 % ethyl acetate:60 % 40:60 petroleum ether). M.P = 101 °C

$v_{max}$/cm$^{-1}$: 2965, 2941, 2879, 1739, 1725, 1682, 1602, 1578, 1510, 1498, 1468, 1450, 1423, 1415, 1397, 1248, 1197, 1187, 1173, 1157, 1119, 1102, 1053, 1030, 1007, 948, 905, 875, 847, 840, 816, 796, 758, 737, 675, 645, 629, 615, 563, 505, 492, 466, 452, 415

$δ_H$/ppm (400 MHz, CDCl$_3$): 10.42 (1 H, s, (C=O)-H), 8.13 (2 H, d, J 8.9 Hz, Ar-H), 7.90 (1 H, d, J 9.0 Hz, Ar-H), 6.98 (2 H, d, J 8.9 Hz, Ar-H), 6.89 (2 H, m, Ar-H), 4.02 (2 H, t, J 6.6 Hz, O-CH$_2$-CH$_2$-), 3.94 (3 H, s, O-CH$_3$), 1.86 (2 H, m, O-CH$_2$-CH$_2$-CH$_3$), 1.07 (3 H, t, J 7.4 Hz, O-CH$_2$-CH$_2$-CH$_3$)

$δ_C$/ppm (100 MHz, CDCl$_3$): 188.83, 164.34, 164.01, 162.98, 157.40, 132.56, 130.04, 122.71, 120.98, 114.58, 114.52, 105.94, 70.00, 56.06, 22.58, 10.61

### *1.4 4-Formyl-3-methoxyphenyl 4-butoxybenzoate*

White solid. Yield: 2.72 g, 55.2 %. RF: 0.279 (100 % dichloromethane). M.P = 89 °C

$v_{max}$/cm$^{-1}$: 2969, 2951, 2909, 2871, 1733, 167, 1601, 1589, 1578, 1511, 1491, 1475, 1461, 1440, 1416, 1400, 1317, 1249, 1200, 1173, 1156, 1121, 1101, 1066, 1051, 1028, 1008, 974, 962, 946, 882, 849, 828, 811, 793, 758, 743, 730, 690, 673, 647, 628, 615, 564, 511, 499, 464, 431, 414

$δ_H$/ppm (400 MHz, CDCl$_3$): 10.42 (1 H, s, (C=O)-H), 8.13 (2 H, d, J 9.0 Hz, Ar-H), 7.90 (1 H, d, J 9.0 Hz, Ar-H), 6.98 (2 H, d, J 9.0 Hz, Ar-H), 6.89 (2 H, m, Ar-H), 4.06 (2 H, t, J 6.5 Hz, O-CH$_2$-CH$_2$-), 3.94 (3 H, s, O-CH$_3$), 1.81 (2 H, m, O-CH$_2$-CH$_2$-CH$_2$-), 1.52 (2 H, m, O-CH$_2$-CH$_2$-CH$_2$-CH$_3$), 1.00 (3 H, t, J 7.4 Hz, O-CH$_2$-CH$_2$-CH$_2$-CH$_3$)

$δ_C$/ppm (100 MHz, CDCl$_3$): 188.69, 164.20, 163.89, 162.85, 157.26, 132.42, 129.90, 122.57, 120.83, 114.44, 114.38, 105.80, 68.09, 55.92, 31.12, 19.20, 13.83

### *1.5 4-Formyl-3-methoxyphenyl 4-pentyloxybenzoate*

White solid. Yield: 2.98 g, 60.4 %. RF: 0.256 (100 % dichloromethane). M.P = 98 °C

$v_{max}$/cm$^{-1}$: 2941, 2873, 1729, 1682, 1601, 1578, 1512, 1499, 1489, 1463, 1416, 1397, 1310, 1248, 1196, 1172, 1154, 1101, 1052, 1028, 1008, 945, 887, 876, 848, 841, 815, 796, 758, 737, 722, 689, 674, 646, 628, 615, 646, 628, 615, 564, 506, 480, 462, 427, 408

$δ_H$/ppm (400 MHz, CDCl$_3$): 10.42 (1 H, s, (C=O)-H), 8.13 (2 H, d, J 9.0 Hz, Ar-H), 7.90 (1 H, d, J 9.0 Hz, Ar-H), 6.98 (2 H, d, J 9.0 Hz, Ar-H), 6.89 (2 H, m, Ar-H), 4.05 (2 H, t, J 6.6 Hz, O-CH$_2$-CH$_2$-), 3.94 (3 H, s, O-CH$_3$), 1.83 (2 H, m, O-CH$_2$-CH$_2$-CH$_2$-), 1.44 (4 H, m, O-CH$_2$-CH$_2$-CH$_2$-CH$_2$-CH$_3$), 0.95 (3 H, t, J 7.1 Hz, O-CH$_2$-CH$_2$-CH$_2$-CH$_2$-CH$_3$)

$δ_C$/ppm (100 MHz, CDCl$_3$): 188.69, 164.20, 163.88, 162.85, 157.26, 132.42, 129.90, 122.58, 120.83, 114.44, 114.38, 105.80, 68.40, 55.92, 28.78, 28.13, 22.44, 14.02

## *4-((4-Alkyloxybenzoyl)oxy)-2-methoxybenzoic acid (2)*

To a pre-dried flask flushed with argon, **Compound 1** (1 eq) and resorcinol (1.5 eq) were solubilised in DMSO (100 mL). Sodium chlorite (4 eq) and sodium hydrogen phosphate monohydrate (3.5 eq) were solubilised in water (60 mL) before being slowly poured into the reaction flask and the resultant mixture was stirred at room temperature overnight. The quantities of the reagents used in each reaction are listed in **Table S2**. The extent of the reaction was monitored by TLC using an appropriate solvent system (RF values quoted in the product data). The reaction mixture was diluted with water (200 mL) and the pH of the mixture was adjusted to 1 using 32% hydrochloric acid (≈ 30 mL). A white solid precipitated after acidification which was collected by vacuum filtration and recrystallised from hot ethanol (250 mL).

**Table S2.** Quantities of reagents used in the syntheses of the 4-((4-alkyloxybenzoyl)oxy)-2-methoxybenzoic acids (**2**)

| $n$ | (**1**) | Sodium Chlorite | Sodium Hydrogen Phosphate Monohydrate | Resorcinol |
|---|---|---|---|---|
| 1 | 4.30 g, 0.0150 mol | 5.43 g, 0.0600 mol | 7.24 g, 0.0525 mol | 2.48 g, 0.0225 mol |
| 2 | 3.40 g, 0.0113 mol | 4.09 g, 0.0452 mol | 5.46 g, 0.0396 mol | 1.87 g, 0.0170 mol |
| 3 | 2.00 g, $6.36 \times 10^{-3}$ mol | 2.31 g, 0.0255 mol | 3.08 g, 0.0223 mol | 1.05 g, $9.54 \times 10^{-3}$ mol |
| 4 | 2.50 g, $7.61 \times 10^{-3}$ mol | 2.76 g, 0.0305 mol | 3.67 g, 0.0266 mol | 1.26 g, 0.0114 mol |
| 5 | 2.80 g, $8.18 \times 10^{-3}$ mol | 2.96 g, 0.0327 mol | 3.95 g, 0.0286 mol | 1.35 g, 0.0123 mol |

### *2.1 4-((4-Methoxybenzoyl)oxy)-2-methoxybenzoic acid*

Yield: 3.95 g, 87.1 %. RF: 0.029 (40 % ethyl acetate:60 % 40:60 petroleum ether). M.P = 204 °C

$v_{max}$/cm$^{-1}$: 3012, 1723, 1700, 1667, 1606, 1579, 1517, 1501, 1461, 1435, 1424, 1401, 1309, 1247, 1189, 1173, 1164, 1134, 1103, 1092, 1061, 1024, 952, 927, 876, 847, 826, 817, 799, 772, 760, 737, 692, 659, 626, 606, 553, 511, 457, 439

$\delta_H$/ppm (400 MHz, DMSO-d$_6$): 12.65 (1 H, s, OH), 8.08 (2 H, d, J 8.8 Hz, Ar-H), 7.74 (1 H, d, J 8.4 Hz, Ar-H), 7.10 (3 H, m, Ar-H), 6.91 (1 H, dd, J 8.4 Hz, 1.9 Hz, Ar-H), 3.88 (3 H, s, O-CH$_3$), 3.82 (3 H, s, O-CH$_3$)

$\delta_C$/ppm (100 MHz, DMSO-d$_6$): 167.05, 164.32, 164.27, 159.90, 154.86, 132.58, 132.31, 121.17, 119.04, 114.77, 114.09, 107.30, 56.58, 56.14

### *2.2 4-((4-Ethoxybenzoyl)oxy)-2-methoxybenzoic acid*

Yield: 2.33 g, 65.2 %. RF: 0.020 (100 % dichloromethane). M.P = 202 °C

$v_{max}$/cm$^{-1}$: 2983, 2944, 1735, 1694, 1666, 1604, 1577, 1509, 1476, 1459, 1433, 1398, 1328, 1301, 1242, 1188, 1161, 1132, 1120, 1101, 1090, 1056, 1026, 1004, 953, 924, 874, 848, 835, 825, 813, 800, 791, 772, 759, 738, 690, 670, 626, 595, 522, 510, 446, 435, 406

$\delta_H$/ppm (400 MHz, DMSO-d$_6$): 12.59 (1 H, s, OH), 7.97 (2 H, d, J 8.7 Hz, Ar-H), 7.73 (1 H, d, J 8.4 Hz, Ar-H), 7.01 (1 H, d, J 2.0 Hz, Ar-H), 6.85 (1 H, dd, J 8.4 Hz, 2.0 Hz, Ar-H), 6.72 (1 H, d, J 2.2 Hz, Ar-H), 6.68 (1 H, dd, J 8.7 Hz, 2.2 Hz, Ar-H), 4.15 (2 H, quart, 7.0 Hz, O-CH$_2$-CH$_3$), 3.91 (3 H, s, O-CH$_3$), 1.36 (3 H, t, 7.0 Hz, O-CH$_2$-CH$_3$)

$\delta_C$/ppm (100 MHz, DMSO-d$_6$): 166.58, 163.80, 163.15, 159.43, 154.40, 132.12, 131.84, 120.52, 118.57, 114.67, 113.62, 106.84, 63.72, 56.12, 14.46

### *2.3 4-((4-Propoxybenzoyl)oxy)-2-methoxybenzoic acid*

Yield: 1.62 g, 77.1 %. RF: 0.027 (100 % dichloromethane). M.P = 137 °C

$v_{max}$/cm$^{-1}$: 2974, 2945, 2878, 1734, 1688, 1664, 1601, 1577, 1507, 1499, 1468, 1421, 1403, 1311, 1289, 1237, 1195, 1168, 1027, 1011, 945, 897, 883, 840, 790, 774, 757, 739, 691, 664, 627, 614, 595, 557, 512, 448, 419, 403

δ$_H$/ppm (400 MHz, DMSO-d$_6$): 12.65 (1 H, s, OH), 8.08 (2 H, d, J 8.7 Hz, Ar-H), 7.74 (1 H, d, J 8.4 Hz, Ar-H), 7.10 (3 H, m, Ar-H), 6.91 (1 H, dd, J 8.4 Hz, 1.8 Hz, Ar-H), 4.05 (2 H, t, J 6.6 Hz, O-CH$_2$-CH$_2$-), 3.82 (3 H, s, O-CH$_3$), 1.77 (2 H, m, O-CH$_2$-CH$_2$-CH$_3$), 1.00 (3 H, t, J 7.4 Hz, O-CH$_2$-CH$_2$-CH$_3$)

δ$_C$/ppm (100 MHz, DMSO-d$_6$): 166.58, 163.79, 163.32, 159.43, 154.40, 132.12, 131.83, 120.51, 118.57, 114.69, 113.62, 106.83, 69.46, 56.11, 21.88, 10.30

### *2.4 4-((4-Butoxybenzoyl)oxy)-2-methoxybenzoic acid*

Yield: 2.30 g, 87.8 %. RF: 0.189 (100 % dichloromethane). M.P = 158 °C

ν$_{max}$/cm$^{-1}$: 2957, 2872, 1728, 1685, 1660, 1601, 1575, 1511, 1471, 1435, 1423, 1404, 1301, 1250, 1196, 1162, 1138, 1125, 1108, 1064, 1025, 1002, 952, 909, 871, 852, 840, 808, 789, 776, 758, 743, 690, 664, 643, 624, 596, 561, 514, 481, 467, 433, 407

δ$_H$/ppm (400 MHz, DMSO-d$_6$): 12.67 (1 H, s, OH), 8.07 (2 H, d, J 9.0 Hz, Ar-H), 7.74 (1 H, d, J 8.4 Hz, Ar-H), 7.11 (3 H, m, Ar-H), 6.91 (1 H, dd, J 8.4 Hz, 2.1 Hz, Ar-H), 4.09 (2 H, t, J 6.5 Hz, O-CH$_2$-CH$_2$-), 3.81 (3 H, s, O-CH$_3$), 1.73 (2 H, m, O-CH$_2$-CH$_2$-CH$_2$-), 1.45 (2 H, m, O-CH$_2$-CH$_2$-CH$_2$-CH$_3$), 0.94 (3 H, t, J 7.4 Hz, O-CH$_2$-CH$_2$-CH$_2$-CH$_3$)

δ$_C$/ppm (100 MHz, DMSO-d$_6$): 166.65, 163.86, 163.37, 159.48, 154.45, 132.17, 131.91, 120.53, 118.58, 114.74, 113.68, 106.86, 67.77, 56.15, 30.59, 18.72, 13.74

### *2.5 4-((4-Pentyloxybenzoyl)oxy)-2-methoxybenzoic acid*

Yield: 2.37 g, 80.8 %. RF: 0.184 (100 % dichloromethane). M.P = 142 °C

ν$_{max}$/cm$^{-1}$: 2945, 2871, 1725, 1685, 1663, 1599, 1575, 1510, 1466, 1433, 1421, 1404, 1300, 1243, 1194, 1159, 1138, 1107, 1096, 1065, 1028, 1007, 947, 874, 852, 842, 812, 778, 758, 740, 690, 664, 642, 625, 595, 559, 511, 466, 429

δ$_H$/ppm (400 MHz, DMSO-d$_6$): 12.67 (1 H, s, OH), 8.07 (2 H, d, J 8.8 Hz, Ar-H), 7.74 (1 H, d, J 8.4 Hz, Ar-H), 7.11 (3 H, m, Ar-H), 6.90 (1 H, dd, J 8.4 Hz, 2.1 Hz, Ar-H), 4.08 (2 H, t, J 6.5 Hz, O-CH$_2$-CH$_2$-), 3.81 (3 H, s, O-CH$_3$), 1.75 (2 H, m, O-CH$_2$-CH$_2$-CH$_2$-), 1.38 (4 H, m, O-CH$_2$-CH$_2$-CH$_2$-CH$_2$-CH$_3$), 0.90 (3 H, t, J 7.0 Hz, O-CH$_2$-CH$_2$-CH$_2$-CH$_2$-CH$_3$)

δ$_C$/ppm (100 MHz, DMSO-d$_6$): 166.65, 163.86, 163.36, 159.48, 154.45, 132.17, 131.91, 120.53, 118.57, 114.74, 113.67, 106.86, 68.05, 56.15, 28.24, 27.67, 21.93, 13.97

## 3-Fluoro-4-nitrophenyl 2-methoxy-4-((4-alkyloxybenzoyl)oxy)benzoates (3)

To a pre-dried flask flushed with argon, **Compound 2** (1 eq), 3-fluoro-4-nitrophenol (1.2 eq), and *N,N'*-dicyclohexylcarbodiimide (1.5 eq) were added to the flask. The solids were solubilised with dichloromethane (30 mL) and stirred for 30 min before 4-dimethylaminopyridine (0.15 eq) was added. The quantities of the reagents used in each reaction are listed in **Table S3**. The temperature of the reaction mixture was increased to room temperature and the reaction was allowed to proceed overnight. For the reactions with *N,N'*-dicyclohexylcarbodiimide, the white precipitate which formed was removed by vacuum filtration and the filtrate collected. The solvent was removed under vacuum and the crude product was purified using a silica gel column with an appropriate solvent system (RF values quoted in product data). The eluent fractions of interest were evaporated under vacuum to leave a white solid which was recrystallised from hot ethanol (60 mL).

**Table S3.** Quantities of reagents used in the syntheses of the 3-fluoro-4-nitrophenyl 2-methoxy-4-((4-alkyloxybenzoyl)oxy)benzoates (**3**)

| $n$ | (**2**) | 3-Fluoro-4-nitrophenol | 4-Dimethylaminopyridine | $N,N'$-Dicyclohexylcarbodiimide |
|---|---|---|---|---|
| 1 | 0.300 g, $9.92 \times 10^{-4}$ mol | 0.187 g, $1.19 \times 10^{-3}$ mol | 18 mg, $1.49 \times 10^{-4}$ mol | 0.307 g, $1.49 \times 10^{-3}$ mol |
| 2 | 0.300 g, $9.48 \times 10^{-4}$ mol | 0.177 g, $1.13 \times 10^{-3}$ mol | 17 mg, $1.42 \times 10^{-4}$ mol | 0.293 g, $1.42 \times 10^{-3}$ mol |
| 3 | 0.300 g, $9.08 \times 10^{-4}$ mol | 0.171 g, $1.09 \times 10^{-3}$ mol | 17 mg, $1.36 \times 10^{-4}$ mol | 0.281 g, $1.36 \times 10^{-3}$ mol |
| 4 | 0.300 g, $8.71 \times 10^{-4}$ mol | 0.164 g, $1.05 \times 10^{-3}$ mol | 16 mg, $1.31 \times 10^{-4}$ mol | 0.270 g, $1.31 \times 10^{-3}$ mol |
| 5 | 0.300 g, $8.37 \times 10^{-4}$ mol | 0.157 g, $1.00 \times 10^{-3}$ mol | 15 mg, $1.26 \times 10^{-4}$ mol | 0.260 g, $1.26 \times 10^{-3}$ mol |

### *3.1 3-Fluoro-4-nitrophenyl 2-methoxy-4-((4-methoxybenzoyl)oxy)benzoate (1OEC3F)*

Yield: 0.285 g, 65.1 %. RF: 0.306 (100 % dichloromethane).

$T_{CrI}$ 178 °C $T_{N_FN}$ (138 °C) $T_{NI}$ (154 °C)

$v_{max}$/cm$^{-1}$: 1763, 1723, 1716, 1604, 1584, 1531, 1513, 1497, 1472, 1454, 1412, 1350, 1320, 1288, 1259, 1249, 1232, 1219, 1193, 1160, 1141, 1119, 1093, 1070, 1056, 1028, 1007, 989, 953, 877, 845, 820, 789, 760, 745, 680, 671, 624, 608, 568, 545, 524, 510, 423, 409

$\delta_H$/ppm (400 MHz, DMSO-d$_6$): 8.30 (1 H, t, J 8.9 Hz, Ar-H), 8.11 (3 H, m, Ar-H), 7.73 (1 H, dd, J 12.1 Hz, 2.3 Hz, Ar-H), 7.43 (1 H, m, Ar-H), 7.26 (1 H, d, J 2.0 Hz, Ar-H), 7.15 (2 H, d, J 8.8 Hz, Ar-H), 7.07 (1 H, dd, J 8.6 Hz, 2.0 Hz, Ar-H), 3.90 (3 H, s, O-CH$_3$), 3.89 (3 H, s, O-CH$_3$)

$\delta_F$/ppm (376 MHz, DMSO-d$_6$): -115.41

$\delta_C$/ppm (100 MHz, DMSO-d$_6$): 164.42, 164.11, 162.14, 161.45, 157.05, 156.69, 156.16, 156.04, 154.43, 135.08, 135.01, 133.82, 132.67, 127.97, 127.95, 121.00, 119.75, 119.71, 115.00, 114.82, 114.61, 113.40, 113.16, 107.80, 56.98, 56.18

MS = [M+Na]$^+$ : Calculated for C$_{22}$H$_{16}$FNO$_8$Na: 464.0758. Found: 464.0777. Difference: 4.1 ppm

### *3.2 3-Fluoro-4-nitrophenyl 2-methoxy-4-((4-ethoxybenzoyl)oxy)benzoate (2OEC3F)*

Yield: 0.087 g, 20.2 %. RF: 0.410 (40 % ethyl acetate:60 % 40:60 petroleum ether).

$T_{CrI}$ 168 °C $T_{N_FN_X}$ (103 °C) $T_{N_XN}$ (118 °C) $T_{NI}$ (153 °C)

$v_{max}$/cm$^{-1}$: 1763, 1723, 1716, 1604, 1584, 1531, 1513, 1497, 1472, 1454, 1412, 1350, 1320, 1288, 1259, 1249, 1232, 1219, 1193, 1160, 1141, 1119, 1093, 1070, 1056, 1028, 1007, 989, 953, 877, 845, 820, 789, 760, 745, 680, 671, 624, 608, 568, 545, 524, 510, 423, 409

$\delta_H$/ppm (400 MHz, CDCl$_3$): 8.15 (4 H, m, Ar-H), 7.28 (1 H, dd, J 11.4 Hz, 2.4 Hz, Ar-H), 7.21 (1 H, m, Ar-H), 6.98 (4 H, m, Ar-H), 4.15 (2 H, quart, 7.0 Hz, O-CH$_2$-CH$_3$), 3.96 (3 H, s, O-CH$_3$), 1.48 (3 H, t, 7.0 Hz, O-CH$_2$-CH$_3$)

δ_F/ppm (376 MHz, CDCl$_3$): -113.38
δ_C/ppm (100 MHz, CDCl$_3$): 164.15, 163.75, 161.88, 161.81, 157.54, 156.86, 155.99, 155.89, 154.89, 134.69, 134.62, 133.78, 132.47, 127.12, 127.10, 120.79, 118.22, 118.18, 114.46, 114.44, 113.94, 112.56, 112.33, 106.50, 63.93, 56.35, 14.67
MS = [M+H]$^+$ : Calculated for C$_{23}$H$_{19}$FNO$_8$: 456.1095. Found: 456.1100. Difference: 1.1 ppm

### 3.3 3-Fluoro-4-nitrophenyl 2-methoxy-4-((4-propoxybenzoyl)oxy)benzoate (3OEC3F)

Yield: 0.240 g, 56.3 %. RF: 0.361 (100 % dichloromethane).
T$_{CrI}$ 145 °C T$_{N_XN}$ (79 °C) T$_{NI}$ (133 °C)
ν$_{max}$/cm$^{-1}$: 3077, 2964, 2881, 1752, 1722, 1601, 1576, 1526, 1508, 1490, 1482, 1423, 1408, 1343, 1283, 1252, 1216, 1190, 1158, 1143, 1129, 1092, 1066, 1042, 1006, 948, 889, 842, 806, 762, 747, 684, 670, 613, 601, 553, 525, 513, 457, 429, 419, 401
δ_H/ppm (400 MHz, CDCl$_3$): 8.15 (4 H, m, Ar-H), 7.28 (1 H, dd, J 11.4 Hz, 2.4 Hz, Ar-H), 7.21 (1 H, m, Ar-H), 6.97 (4 H, m, Ar-H), 4.03 (2 H, t, J 6.6 Hz, O-CH$_2$-CH$_2$-), 3.96 (3 H, s, O-CH$_3$), 1.87 (2 H, m, O-CH$_2$-CH$_2$-CH$_3$), 1.07 (3 H, t, J 7.4 Hz, O-CH$_2$-CH$_2$-CH$_3$)
δ_F/ppm (376 MHz, CDCl$_3$): -113.35
δ_C/ppm (100 MHz, CDCl$_3$): 163.99, 163.75, 161.67, 161.63, 157.34, 156.67, 155.79, 155.69, 154.69, 134.47, 134.40, 133.60, 132.27, 126.94, 126.92, 120.49, 118.05, 118.01, 114.28, 114.18, 113.76, 112.39, 112.15, 106.30, 69.68, 56.16, 22.26, 10.30
MS = [M+H]$^+$ : Calculated for C$_{24}$H$_{21}$FNO$_8$: 470.1251. Found: 470.1268. Difference: 3.6 ppm

### 3.4 3-Fluoro-4-nitrophenyl 2-methoxy-4-((4-butoxybenzoyl)oxy)benzoate (4OEC3F)

Yield: 0.242 g, 55.1 %. RF: 0.351 (100 % dichloromethane).
T$_{CrN}$ 128 °C T$_{N_XN}$ (38 °C) T$_{NI}$ 130 °C
ν$_{max}$/cm$^{-1}$: 2961, 2876, 1726, 1715, 1602, 1577, 1530, 1511, 1493, 1474, 1415, 1348, 1318, 1260, 1227, 1189, 1163, 1126, 1116, 1092, 1073, 1047, 1021, 968, 947, 885, 842, 820, 760, 744, 691, 672, 629, 616, 573, 546, 511, 461, 420, 403
δ_H/ppm (400 MHz, CDCl$_3$): 8.14 (4 H, m, Ar-H), 7.28 (1 H, dd, J 11.4 Hz, 2.4 Hz, Ar-H), 7.21 (1 H, m, Ar-H), 6.97 (4 H, m, Ar-H), 4.07 (2 H, t, J 6.5 Hz, O-CH$_2$-CH$_2$-), 3.96 (3 H, s, O-CH$_3$), 1.82 (2 H, m, O-CH$_2$-CH$_2$-CH$_2$-), 1.53 (2 H, m, O-CH$_2$-CH$_2$-CH$_2$-CH$_3$), 1.00 (3 H, t, J 7.4 Hz, O-CH$_2$-CH$_2$-CH$_2$-CH$_3$)
δ_F/ppm (376 MHz, CDCl$_3$): -113.35
δ_C/ppm (100 MHz, CDCl$_3$): 164.30, 164.08, 161.98, 161.94, 157.66, 156.98, 156.11, 156.00, 155.01, 134.78, 134.71, 133.91, 132.58, 127.26, 127.24, 120.80, 118.36, 118.32, 114.59, 114.49, 114.07, 112.70, 112.47, 106.61, 68.23, 56.48, 31.24, 19.33, 13.97
MS = [M+H]$^+$ : Calculated for C$_{25}$H$_{23}$FNO$_8$: 484.1408. Found: 484.1418. Difference: 2.1 ppm

### 3.5 3-Fluoro-4-nitrophenyl 2-methoxy-4-((4-pentyloxybenzoyl)oxy)benzoate (5OEC3F)

Yield: 0.176 g, 42.3 %. RF: 0.417 (100 % dichloromethane).
T$_{CrN}$ 92 °C T$_{NI}$ 118 °C
ν$_{max}$/cm$^{-1}$: 2935, 2871, 1761, 1718, 1601, 1578, 1529, 1510, 1486, 1474, 1414, 1346, 1318, 1259, 1223, 1192, 1152, 1124, 1115, 1093, 1079, 1043, 1023, 1005, 969, 886, 841, 818, 759, 744, 691, 672, 628, 614, 573, 545, 513, 462, 427, 415
δ_H/ppm (400 MHz, CDCl$_3$): 8.15 (4 H, m, Ar-H), 7.28 (1 H, dd, J 11.3 Hz, 2.4 Hz, Ar-H), 7.21 (1 H, m, Ar-H), 6.97 (4 H, m, Ar-H), 4.06 (2 H, t, J 6.6 Hz, O-CH$_2$-CH$_2$-), 3.96 (3 H, s, O-CH$_3$), 1.83 (2 H, m, O-CH$_2$-CH$_2$-CH$_2$-), 1.44 (4 H, m, O-CH$_2$-CH$_2$-CH$_2$-CH$_2$-CH$_3$), 0.95 (3 H, t, J 7.1 Hz, O-CH$_2$-CH$_2$-CH$_2$-CH$_2$-CH$_3$)

δ_F/ppm (376 MHz, CDCl$_3$): -113.37

δ_C/ppm (100 MHz, CDCl$_3$): 164.31, 164.08, 162.00, 161.95, 157.67, 156.99, 156.12, 156.01, 155.02, 134.80, 134.73, 133.92, 132.59, 127.27, 127.24, 120.81, 118.37, 118.33, 114.60, 114.51, 114.08, 112.71, 112.47, 106.62, 68.55, 56.49, 28.91, 28.27, 22.58, 14.17

MS = [M+H]$^+$ : Calculated for C$_{26}$H$_{25}$FNO$_8$: 498.1564. Found: 498.1582. Difference: 3.6 ppm

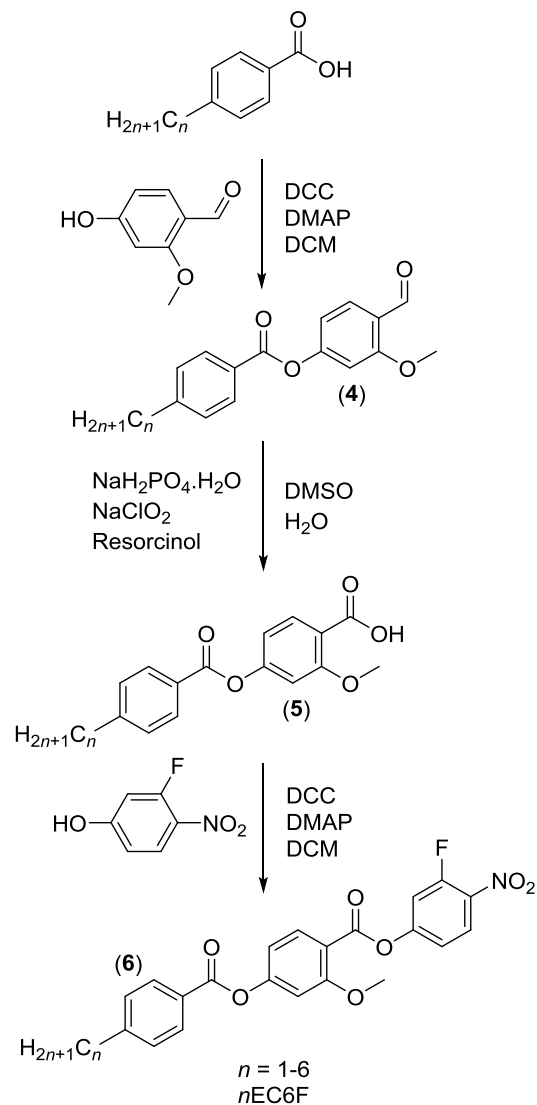

### *4-Formyl-3-methoxyphenyl 4-alkylbenzoates (4)*

To a pre-dried flask flushed with argon, 4-alkylbenzoic acid of the appropriate chain length (1 eq), 4-hydroxy-2-methoxybenzaldehyde (1.1 eq) and 4-dimethylaminopyridine (0.13 eq) were added. The solids were solubilised with dichloromethane (100 mL) and tetrahydrofuran (20 mL) while being stirred for 10 min before *N,N'*-dicyclohexylcarbodiimide (1.3 eq) was added to the flask and the reaction was allowed to proceed overnight. The quantities of the reagents used in each reaction are listed in **Table S4**. The extent of the reaction was monitored by TLC using an appropriate solvent system (RF values quoted in the product data). The precipitate which formed was removed by vacuum filtration and the filtrate collected. The collected solvent was evaporated under vacuum to leave a solid which was recrystallised from hot ethanol (150 mL).

**Table S4.** Quantities of reagents used in the syntheses of the 4-formyl-3-methoxyphenyl 4-alkylbenzoates (**4**)

| $n$ | 4-Alkylbenzoic Acid | 4-Hydroxy-2-methoxybenzaldehyde | 4-Dimethylaminopyridine | $N,N'$-Dicyclohexylcarbodiimide |
|---|---|---|---|---|
| 1 | 2.50 g, 0.0184 mol | 3.07 g, 0.0202 mol | 0.292 g, 2.39×10$^{-3}$ mol | 4.93 g, 0.0256 mol |
| 2 | 3.00 g, 0.0200 mol | 3.34 g, 0.0220 mol | 0.318 g, 2.60×10$^{-3}$ mol | 5.36 g, 0.0260 mol |
| 3 | 3.00 g, 0.0183 mol | 3.06 g, 0.0201 mol | 0.291 g, 2.38×10$^{-3}$ mol | 4.91 g, 0.0238 mol |
| 4 | 3.00 g, 0.0168 mol | 2.81 g, 0.0185 mol | 0.266 g, 2.18×10$^{-3}$ mol | 4.50 g, 0.0218 mol |
| 5 | 3.00 g, 0.0156 mol | 2.62 g, 0.0172 mol | 0.248 g, 2.03×10$^{-3}$ mol | 4.19 g, 0.0203 mol |
| 6 | 3.00 g, 0.0145 mol | 2.43 g, 0.0160 mol | 0.231 g, 1.89×10$^{-3}$ mol | 3.90 g, 0.0189 mol |

*4.1 4-Formyl-3-methoxyphenyl 4-methylbenzoate*

White solid. Yield: 3.35 g, 67.4 %. RF: 0.550 (40 % ethyl acetate:60 % 40:60 petroleum ether). M.P = 117 °C

$v_{max}$/cm$^{-1}$: 2869, 1739, 1679, 1599, 1587, 1493, 1474, 1465, 1417, 1403, 1379, 1262, 1248, 1200, 1182, 1157, 1120, 1099, 1063, 1027, 1017, 945, 872, 838, 822, 806, 788, 743, 686, 668, 627, 604, 561, 502, 467, 435, 409

$\delta_H$/ppm (400 MHz, CDCl$_3$): 10.42 (1 H, s, (C=O)-H), 8.08 (2 H, d, J 8.3 Hz, Ar-H), 7.91 (1 H, d, J 7.9 Hz, Ar-H), 7.33 (2 H, d, J 8.3 Hz, Ar-H), 6.90 (2 H, m, Ar-H), 3.94 (3 H, s, O-<u>CH$_3$</u>), 2.46 (3 H, s, Ar-<u>CH$_3$</u>)

$\delta_C$/ppm (100 MHz, CDCl$_3$): 188.80, 164.64, 162.98, 157.27, 145.13, 130.44, 130.06, 129.58, 126.29, 122.79, 114.47, 105.90, 56.06, 21.95

*4.2 4-Formyl-3-methoxyphenyl 4-ethylbenzoate*

White solid. Yield: 2.99 g, 52.6 %. RF: 0.600 (40 % ethyl acetate:60 % 40:60 petroleum ether). M.P = 105 °C

$v_{max}$/cm$^{-1}$: 2971, 2880, 1732, 1680, 1600, 1583, 1488, 1470, 1457, 1416, 1399, 1305, 1252, 1191, 1179, 1151, 1100, 1065, 1051, 1042, 1024, 943, 888, 851, 816, 803, 777, 728, 695, 663, 636, 627, 599, 563, 497, 465, 455, 434, 422

$\delta_H$/ppm (400 MHz, CDCl$_3$): 10.43 (1 H, s, (C=O)-H), 8.11 (2 H, d, J 8.2 Hz, Ar-H), 7.91 (1 H, d, J 8.9 Hz, Ar-H), 7.35 (2 H, d, J 8.2 Hz, Ar-H), 6.90 (2 H, m, Ar-H), 3.94 (3 H, s, O-<u>CH$_3$</u>), 2.76 (2 H, quart, 7.6 Hz, Ar-<u>CH$_2$</u>-CH$_3$), 1.29 (3 H, t, 7.6 Hz, Ar-CH$_2$-<u>CH$_3$</u>)

$\delta_C$/ppm (100 MHz, CDCl$_3$): 188.79, 164.64, 162.98, 157.28, 151.29, 130.56, 130.06, 128.40, 126.48, 122.79, 114.47, 105.90, 56.06, 29.22, 15.34

### *4.3 4-Formyl-3-methoxyphenyl 4-propylbenzoate*

Off-white solid. Yield: 4.28 g, 78.4 %. RF: 0.583 (40 % ethyl acetate:60 % 40:60 petroleum ether). M.P = 86 °C

$v_{max}$/cm$^{-1}$: 2955, 2928, 2859, 1739, 1680, 1599, 1582, 1487, 1464, 1415, 1393, 1311, 1270, 1239, 1193, 1174, 1149, 1100, 1072, 1054, 1015, 941, 901, 887, 851, 810, 793, 755, 737, 696, 670, 643, 634, 610, 560, 514, 500, 464, 449, 418, 406

$\delta_H$/ppm (400 MHz, CDCl$_3$): 10.42 (1 H, s, (C=O)-H), 8.10 (2 H, d, J 8.2 Hz, Ar-H), 7.91 (1 H, d, J 9.0 Hz, Ar-H), 7.33 (2 H, d, J 8.2 Hz, Ar-H), 6.90 (2 H, m, Ar-H), 3.94 (3 H, s, O-CH$_3$), 2.69 (2 H, t, J 7.4 Hz, Ar-CH$_2$-CH$_2$-), 1.70 (2 H, m, Ar-CH$_2$-CH$_2$-CH$_3$), 0.97 (3 H, t, J 7.4 Hz, Ar-CH$_2$-CH$_2$-CH$_3$)

$\delta_C$/ppm (100 MHz, CDCl$_3$): 188.81, 164.66, 162.98, 157.29, 149.79, 130.46, 130.06, 128.99, 126.50, 122.78, 114.47, 105.90, 56.06, 38.27, 24.36, 13.88

### *4.4 4-Formyl-3-methoxyphenyl 4-butylbenzoate*

Off-white solid. Yield: 2.64 g, 50.4 %. RF: 0.342 (100 % dichloromethane). M.P = 61 °C

$v_{max}$/cm$^{-1}$: 2951, 2929, 2868, 1739, 1673, 1603, 1580, 1486, 1470, 1448, 1414, 1404, 1373, 1308, 1273, 120, 1239, 1207, 1171, 1150, 1106, 1075, 1056, 1028, 1017, 944, 902, 870, 841, 804, 753, 731, 693, 657, 635, 621, 570, 520, 489, 46, 410, 403

$\delta_H$/ppm (400 MHz, CDCl$_3$): 10.42 (1 H, s, (C=O)-H), 8.10 (2 H, d, J 8.4 Hz, Ar-H), 7.91 (1 H, d, J 8.9 Hz, Ar-H), 7.33 (2 H, d, J 8.4 Hz, Ar-H), 6.90 (2 H, m, Ar-H), 3.94 (3 H, s, O-CH$_3$), 2.71 (2 H, t, J 7.6 Hz, Ar-CH$_2$-CH$_2$-), 1.64 (2 H, m, Ar-CH$_2$-CH$_2$-CH$_2$-), 1.38 (2 H, m, Ar-CH$_2$-CH$_2$-CH$_2$-CH$_3$), 0.95 (3 H, t, J 7.3 Hz, Ar-CH$_2$-CH$_2$-CH$_2$-CH$_3$)

$\delta_C$/ppm (100 MHz, CDCl$_3$): 188.84, 164.66, 162.97, 157.28, 150.06, 130.47, 130.06, 128.94, 126.41, 122.74, 114.47, 105.89, 56.05, 35.95, 33.38, 22.45, 14.06

### *4.5 4-Formyl-3-methoxyphenyl 4-pentylbenzoate*

Off-white solid. Yield: 2.22 g, 43.6 %. RF: 0.367 (100 % dichloromethane). M.P = 73 °C

$v_{max}$/cm$^{-1}$: 2955, 2928, 2859, 2780, 1738, 1672, 1602, 1487, 1469, 1449, 1416, 1405, 1308, 1273, 1251, 1206, 1171, 1103, 1077, 1058, 1027, 1016, 944, 891, 870, 840, 808, 752, 727, 694, 658, 634, 623, 611, 571, 528, 488, 467, 441, 432, 413

$\delta_H$/ppm (400 MHz, CDCl$_3$): 10.42 (1 H, s, (C=O)-H), 8.10 (2 H, d, J 8.1 Hz, Ar-H), 7.91 (1 H, d, J 8.9 Hz, Ar-H), 7.33 (2 H, d, J 8.1 Hz, Ar-H), 6.90 (2 H, m, Ar-H), 3.94 (3 H, s, O-CH$_3$), 2.70 (2 H, t, J 7.8 Hz, Ar-CH$_2$-CH$_2$-), 1.67 (2 H, tt, J 7.8 Hz, 7.2 Hz, Ar-CH$_2$-CH$_2$-CH$_2$-), 1.34 (4 H, m, Ar-CH$_2$-CH$_2$-CH$_2$-CH$_2$-CH$_3$), 0.90 (3 H, t, J 7.0 Hz, Ar-CH$_2$-CH$_2$-CH$_2$-CH$_2$-CH$_3$)

$\delta_C$/ppm (100 MHz, CDCl$_3$): 188.83, 164.66, 162.97, 157.28, 150.08, 130.47, 130.05, 128.93, 126.40, 122.74, 114.46, 105.88, 56.04, 36.21, 31.54, 30.93, 22.63, 14.14

### *4.6 4-Formyl-3-methoxyphenyl 4-hexylbenzoate*

Off-white solid. Yield: 2.52 g, 51.1 %. RF: 0.432 (100 % dichloromethane). M.P = 69 °C

$v_{max}$/cm$^{-1}$: 2953, 2926, 2854, 2781, 1735, 1677, 1599, 1586, 1487, 1466, 1416, 1404, 1310, 1244, 1201, 1175, 1152, 1100, 1058, 1028, 1016, 945, 890, 841, 817, 796, 753, 726, 697, 669, 645, 634, 610, 564, 521, 496, 465, 408

$\delta_H$/ppm (400 MHz, CDCl$_3$): 10.42 (1 H, s, (C=O)-H), 8.10 (2 H, d, J 8.3 Hz, Ar-H), 7.91 (1 H, d, J 9.0 Hz, Ar-H), 7.33 (2 H, d, J 8.3 Hz, Ar-H), 6.90 (2 H, m, Ar-H), 3.94 (3 H, s, O-CH$_3$), 2.71 (2 H, t, J 7.6 Hz, Ar-CH$_2$-CH$_2$-), 1.88 (2 H, tt, J 7.6 Hz, 6.9 Hz, Ar-CH$_2$-CH$_2$-CH$_2$-), 1.32 (6 H, m, Ar-CH$_2$-CH$_2$-CH$_2$-CH$_2$-CH$_2$-CH$_3$), 0.89 (3 H, t, J 6.9 Hz, Ar-CH$_2$-CH$_2$-CH$_2$-CH$_2$-CH$_2$-CH$_3$)

δ$_C$/ppm (100 MHz, CDCl$_3$): 188.86, 164.67, 162.98, 157.29, 150.10, 130.48, 130.07, 128.94, 126.40, 122.74, 114.48, 105.89, 56.05, 36.26, 31.79, 31.22, 29.04, 22.72, 14.23

## 4-((4-Alkylbenzoyl)oxy)-2-methoxybenzoic acid (5)

To a pre-dried flask flushed with argon, **Compound 4** (1 eq) and resorcinol (1.5 eq) were solubilised in DMSO (100 mL or 120 mL for *n* = 3). Sodium chlorite (4 eq) and sodium hydrogen phosphate monohydrate (3.5 eq) were solubilised in water (60 mL or 80 mL for *n* = 3) before being slowly poured into the reaction flask and the resultant mixture was stirred at room temperature overnight. The quantities of the reagents used in each reaction are listed in **Table S5**. The extent of the reaction was monitored by TLC using an appropriate solvent system (RF values quoted in the product data). The reaction mixture was diluted with water (200 mL) and the pH of the mixture was adjusted to 1 using 32% hydrochloric acid (≈ 30 mL). A solid precipitated after acidification which was collected by vacuum filtration and recrystallised from hot ethanol (150 mL) or hot ethanol (70 mL) with 40:60 petroleum ether (60 mL) for *n* = 6.

**Table S5.** Quantities of reagents used in the syntheses of the 4-((4-alkyloxybenzoyl)oxy)-2-methoxybenzoic acids (**5**)

| *n* | (**4**) | Sodium Chlorite | Sodium Hydrogen Phosphate Monohydrate | Resorcinol |
|---|---|---|---|---|
| 1 | 3.10 g, 0.0115 mol | 4.16 g, 0.0460 mol | 5.56 g, 0.0403 mol | 1.90 g, 0.0173 mol |
| 2 | 2.80 g, 9.85×10$^{-3}$ mol | 3.56 g, 0.0394 mol | 4.76 g, 0.0345 mol | 1.63 g, 0.0148 mol |
| 3 | 4.00 g, 0.0134 mol | 4.85 g, 0.0536 mol | 6.47 g, 0.0469 mol | 2.21 g, 0.0201 mol |
| 4 | 2.50 g, 8.00×10$^{-3}$ mol | 2.89 g, 0.0320 mol | 3.86 g, 0.0280 mol | 1.32 g, 0.0120 mol |
| 5 | 2.10 g, 6.43×10$^{-3}$ mol | 2.32 g, 0.0257 mol | 3.10 g, 0.0225 mol | 1.06 g, 9.65×10$^{-3}$ mol |
| 6 | 2.40 g, 7.05×10$^{-3}$ mol | 2.55 g, 0.0282 mol | 3.41 g, 0.0247 mol | 1.17 g, 0.0106 mol |

### 5.1 4-((4-Methylbenzoyl)oxy)-2-methoxybenzoic acid

Off-white solid. Yield: 2.54 g, 77.2 %. RF: 0.079 (40 % ethyl acetate:60 % 40:60 petroleum ether). M.P = 204 °C

ν$_{max}$/cm$^{-1}$: 2816, 1725, 1685, 1672, 1604, 1583, 1499, 1468, 1408, 1303, 1241, 1194, 1177, 1160, 1140, 1095, 1060, 1029, 1018, 947, 892, 838, 787, 770, 747, 668, 652, 593, 554, 481, 445

δ$_H$/ppm (400 MHz, DMSO-d$_6$): 12.67 (1 H, s, OH), 8.03 (2 H, d, J 8.0 Hz, Ar-H), 7.75 (1 H, d, J 8.4 Hz, Ar-H), 7.42 (2 H, d, J 8.0 Hz, Ar-H), 7.11 (1 H, d, J 2.0 Hz, Ar-H), 6.92 (1 H, dd, J 8.4 Hz, 2.0 Hz, Ar-H), 3.82 (3 H, s, O-CH$_3$), 2.43 (3 H, s, Ar-CH$_3$)

δ$_C$/ppm (100 MHz, DMSO-d$_6$): 166.58, 164.16, 159.45, 154.32, 144.72, 131.87, 129.93, 129.56, 125.93, 118.70, 113.58, 106.82, 56.13, 21.28

### 5.2 4-((4-Ethylbenzoyl)oxy)-2-methoxybenzoic acid

White solid. Yield: 1.73 g, 58.5 %. RF: 0.105 (40 % ethyl acetate:60 % 40:60 petroleum ether). M.P = 170 °C

$v_{max}$/cm$^{-1}$: 2979, 2941, 1725, 1683, 1669, 1604, 1580, 1496, 1464, 1433, 1404, 1285, 1238, 1190, 1177, 1157, 1138, 1093, 1065, 1051, 1026, 1016, 979, 947, 889, 850, 781, 766, 734, 696, 650, 627, 594, 555, 496, 442, 418

$δ_H$/ppm (400 MHz, DMSO-d$_6$): 12.67 (1 H, s, OH), 8.06 (2 H, d, J 8.1 Hz, Ar-H), 7.75 (1 H, d, J 8.4 Hz, Ar-H), 7.45 (2 H, d, J 8.0 Hz, Ar-H), 7.11 (1 H, d, J 2.0 Hz, Ar-H), 6.92 (1 H, dd, J 8.4 Hz, 2.0 Hz, Ar-H), 3.82 (3 H, s, O-CH$_3$), 2.73 (2 H, quart, 7.6 Hz, Ar-CH$_2$-CH$_3$), 1.22 (3 H, t, 7.6 Hz, Ar-CH$_2$-CH$_3$)

$δ_C$/ppm (100 MHz, DMSO-d$_6$): 166.58, 164.16, 159.44, 154.32, 150.75, 131.87, 130.05, 128.39, 126.17, 118.70, 113.59, 106.82, 56.13, 28.27, 15.18

### 5.3 4-((4-Propylbenzoyl)oxy)-2-methoxybenzoic acid

White solid. Yield: 2.58 g, 61.3 %. RF: 0.054 (40 % ethyl acetate:60 % 40:60 petroleum ether). M.P = 166 °C

$v_{max}$/cm$^{-1}$: 2961, 2871, 1731, 1690, 1666, 1600, 1581, 1496, 1468, 1406, 1297, 1237, 1194, 1177, 1163, 1130, 1091, 1056, 1028, 1016, 940, 891, 851, 840, 788, 774, 758, 731, 692, 664, 634, 616, 594, 553, 491, 462, 441, 416

$δ_H$/ppm (400 MHz, DMSO-d$_6$): 12.66 (1 H, s, OH), 8.06 (2 H, d, J 8.2 Hz, Ar-H), 7.75 (1 H, d, J 8.4 Hz, Ar-H), 7.43 (2 H, d, J 8.2 Hz, Ar-H), 7.11 (1 H, d, J 2.0 Hz, Ar-H), 6.92 (1 H, dd, J 8.4 Hz, 2.0 Hz, Ar-H), 3.82 (3 H, s, O-CH$_3$), 2.68 (2 H, t, J 7.6 Hz, Ar-CH$_2$-CH$_2$-), 1.64 (2 H, m, Ar-CH$_2$-CH$_2$-CH$_3$), 0.90 (3 H, t, J 7.3 Hz, Ar-CH$_2$-CH$_2$-CH$_3$)

$δ_C$/ppm (100 MHz, DMSO-d$_6$): 166.57, 164.15, 159.44, 154.33, 149.14, 131.86, 129.95, 128.97, 126.21, 118.69, 113.59, 106.82, 56.12, 37.17, 23.73, 13.52

### 5.4 4-((4-Butylbenzoyl)oxy)-2-methoxybenzoic acid

White solid. Yield: 1.13 g, 43.0 %. RF: 0.162 (100 % dichloromethane). M.P = 115 °C

$v_{max}$/cm$^{-1}$: 2955, 2932, 2872, 1727, 1699, 1677, 1603, 1582, 1495, 1466, 1453, 1429, 1406, 1245, 1190, 1177, 1158, 1131, 1103, 1093, 1061, 1028, 1015, 948, 934, 889, 847, 786, 772, 750, 725, 687, 664, 635, 610, 592, 554, 502, 468, 439, 416, 404

$δ_H$/ppm (400 MHz, DMSO-d$_6$): 12.68 (1 H, s, OH), 8.05 (2 H, d, J 8.3 Hz, Ar-H), 7.75 (1 H, d, J 8.4 Hz, Ar-H), 7.43 (2 H, d, J 8.3 Hz, Ar-H), 7.11 (1 H, d, J 2.1 Hz, Ar-H), 6.92 (1 H, dd, J 8.4 Hz, 2.1 Hz, Ar-H), 3.81 (3 H, s, O-CH$_3$), 2.70 (2 H, t, J 7.6 Hz, Ar-CH$_2$-CH$_2$-), 1.59 (2 H, m, Ar-CH$_2$-CH$_2$-CH$_2$-), 1.31 (2 H, m, Ar-CH$_2$-CH$_2$-CH$_2$-CH$_3$), 0.90 (3 H, t, J 7.3 Hz, Ar-CH$_2$-CH$_2$-CH$_2$-CH$_3$)

$δ_C$/ppm (100 MHz, DMSO-d$_6$): 166.65, 164.21, 159.49, 154.38, 149.44, 131.93, 130.03, 128.97, 126.19, 118.70, 113.65, 106.86, 56.15, 34.88, 32.78, 21.74, 13.80

### 5.5 4-((4-Pentylbenzoyl)oxy)-2-methoxybenzoic acid

Off-white solid. Yield: 1.30 g, 59.0 %. RF: 0.135 (100 % dichloromethane).

T$_{CrI}$ 124 °C T$_{NI}$ (73 °C)

$v_{max}$/cm$^{-1}$: 2947, 2863, 1740, 1663, 1605, 1577, 1500, 1466, 1454, 1430, 1403, 1294, 1240, 1187, 1178, 1158, 1135, 1106, 1056, 1029 1015, 941, 870, 834, 789, 774, 752, 728, 696, 663, 640, 631, 619, 598, 558, 518, 489, 462, 427, 405

$δ_H$/ppm (400 MHz, DMSO-d$_6$): 12.68 (1 H, s, OH), 8.05 (2 H, d, J 8.4 Hz, Ar-H), 7.75 (1 H, d, J 8.4 Hz, Ar-H), 7.43 (2 H, d, J 8.4 Hz, Ar-H), 7.11 (1 H, d, J 2.1 Hz, Ar-H), 6.92 (1 H, dd, J 8.4 Hz, 2.1 Hz, Ar-H), 3.81 (3 H, s, O-CH$_3$), 2.69 (2 H, t, J 7.6 Hz, Ar-CH$_2$-CH$_2$-), 1.61 (2 H, tt, J 7.6 Hz, 7.3 Hz, Ar-CH$_2$-CH$_2$-CH$_2$-), 1.29 (4 H, m, Ar-CH$_2$-CH$_2$-CH$_2$-CH$_2$-CH$_3$), 0.86 (3 H, t, J 7.0 Hz, Ar-CH$_2$-CH$_2$-CH$_2$-CH$_2$-CH$_3$)

δ$_C$/ppm (100 MHz, DMSO-d$_6$): 166.65, 164.22, 159.49, 154.38, 149.46, 131.93, 130.03, 128.98, 126.20, 118.71, 113.65, 106.86, 56.16, 35.15, 30.84, 30.31, 21.97, 13.97

### 5.6 4-((4-Hexylbenzoyl)oxy)-2-methoxybenzoic acid

Brown solid. Yield: 1.46 g, 58.1 %. RF: 0.206 (100 % dichloromethane).

T$_{CrI}$ 106 °C T$_{NI}$ (63 °C)

$v_{max}$/cm$^{-1}$: 2949, 2925, 2852, 1730, 1679, 1603, 1582, 1494, 1466, 1414, 1323, 1299, 1236, 1190, 1174, 1158, 1141, 1096, 1058, 1031, 1017, 949, 891, 875, 848, 828, 791, 773, 759, 739, 726, 696, 663, 628, 596, 554, 526, 463, 443, 407

δ$_H$/ppm (400 MHz, DMSO-d$_6$): 12.72 (1 H, s, OH), 8.05 (2 H, d, J 8.3 Hz, Ar-H), 7.75 (1 H, d, J 8.4 Hz, Ar-H), 7.44 (2 H, d, J 8.3 Hz, Ar-H), 7.12 (1 H, d, J 2.1 Hz, Ar-H), 6.93 (1 H, dd, J 8.4 Hz, 2.1 Hz, Ar-H), 3.82 (3 H, s, O-CH$_3$), 2.70 (2 H, t, J 7.6 Hz, Ar-CH$_2$-CH$_2$-), 1.61 (2 H, tt, J 7.6 Hz, 7.0 Hz, Ar-CH$_2$-CH$_2$-CH$_2$-), 1.28 (6 H, m, Ar-CH$_2$-CH$_2$-CH$_2$-CH$_2$-CH$_2$-CH$_3$), 0.86 (3 H, t, J 7.0 Hz, Ar-CH$_2$-CH$_2$-CH$_2$-CH$_2$-CH$_2$-CH$_3$)

δ$_C$/ppm (100 MHz, DMSO-d$_6$): 166.66, 164.23, 159.49, 154.39, 149.47, 131.94, 130.04, 128.98, 126.20, 118.70, 113.66, 106.86, 56.16, 35.20, 31.11, 30.60, 28.29, 22.11, 14.01

## 3-Fluoro-4-nitrophenyl 2-methoxy-4-((4-alkylbenzoyl)oxy)benzoates (6)

To a pre-dried flask flushed with argon, **Compound 5** (1 eq), 3-fluoro-4-nitrophenol (1.2 eq), and *N,N'*-dicyclohexylcarbodiimide (1.5 eq) were added to the flask. The solids were solubilised with dichloromethane (30 mL) and stirred for 30 min before 4-dimethylaminopyridine (0.15 eq) was added. The quantities of the reagents used in each reaction are listed in **Table S6**. The temperature of the reaction mixture was increased to room temperature and the reaction was allowed to proceed overnight. For the reactions with *N,N'*-dicyclohexylcarbodiimide, the white precipitate which formed was removed by vacuum filtration and the filtrate collected. The solvent was removed under vacuum and the crude product was purified using a silica gel column with an appropriate solvent system (RF values quoted in product data). The eluent fractions of interest were evaporated under vacuum to leave a white solid which was recrystallised from hot ethanol (80 mL).

**Table S6.** Quantities of reagents used in the syntheses of the 3-fluoro-4-nitrophenyl 2-methoxy-4-((4-alkylbenzoyl)oxy)benzoates (**6**)

| n | (5) | 3-Fluoro-4-nitrophenol | 4-Dimethylaminopyridine | N,N'-Dicyclohexylcarbodiimide |
|---|---|---|---|---|
| 1 | 0.300 g, 1.06×10$^{-3}$ mol | 0.198 g, 1.26×10$^{-3}$ mol | 19 mg, 1.58×10$^{-4}$ mol | 0.326 g, 1.58×10$^{-3}$ mol |
| 2 | 0.300 g, 9.99×10$^{-4}$ mol | 0.189 g, 1.20×10$^{-3}$ mol | 18 mg, 1.50×10$^{-4}$ mol | 0.309 g, 1.50×10$^{-3}$ mol |
| 3 | 0.300 g, 9.54×10$^{-4}$ mol | 0.179 g, 1.14×10$^{-3}$ mol | 17 mg, 1.43×10$^{-4}$ mol | 0.295 g, 1.43×10$^{-3}$ mol |

| 4 | 0.300 g, 9.14×10$^{-4}$ mol | 0.173 g, 1.10×10$^{-3}$ mol | 16 mg, 1.37×10$^{-4}$ mol | 0.283 g, 1.37×10$^{-3}$ mol |
|---|---|---|---|---|
| 5 | 0.300 g, 8.76×10$^{-4}$ mol | 0.165 g, 1.05×10$^{-3}$ mol | 16 mg, 1.31×10$^{-4}$ mol | 0.270 g, 1.31×10$^{-3}$ mol |
| 6 | 0.300 g, 8.42×10$^{-4}$ mol | 0.159 g, 1.00×10$^{-3}$ mol | 15 mg, 1.26×10$^{-4}$ mol | 0.260 g, 1.26×10$^{-3}$ mol |

### 6.1 3-Fluoro-4-nitrophenyl 2-methoxy-4-((4-methylbenzoyl)oxy)benzoate (1EC6F)
Yield: 0.111 g, 24.6 %. RF: 0.216 (100 % dichloromethane).
T$_{CrI}$ 169 °C T$_{N_FI}$ (156 °C)
$v_{max}$/cm$^{-1}$: 1754, 1738, 1731, 1610, 1581, 1529, 1484, 1457, 1412, 1347, 1316, 1263, 1219, 1194, 1180, 1168, 1134, 1092, 1069, 1033, 1007, 966, 908, 889, 878, 856, 837, 811, 793, 758, 744, 734, 684, 637, 614, 556, 529, 476, 455
$\delta_H$/ppm (400 MHz, CDCl$_3$): 8.14 (4 H, m, Ar-H), 7.34 (2 H, d, J 8.2 Hz, Ar-H), 7.29 (1 H, dd, J 11.4 Hz, 2.4 Hz, Ar-H), 7.21 (1 H, m, Ar-H), 6.97 (2 H, m, Ar-H), 3.97 (3 H, s, O-CH$_3$), 2.48 (3 H, s, Ar-CH$_3$)
$\delta_F$/ppm (376 MHz, CDCl$_3$): -113.37
$\delta_C$/ppm (100 MHz, CDCl$_3$): 164.61, 162.00, 161.95, 157.68, 156.88, 156.11, 156.01, 155.03, 145.26, 134.84, 134.77, 133.94, 130.48, 129.62, 127.26, 127.25, 126.19, 118.35, 118.31, 114.70, 114.04, 112.70, 112.46, 106.60, 56.51, 21.98
MS = [M+Na]$^+$ : Calculated for C$_{22}$H$_{16}$FNO$_7$Na: 448.0808. Found: 448.0814. Difference: 1.3 ppm

### 6.2 3-Fluoro-4-nitrophenyl 2-methoxy-4-((4-ethylbenzoyl)oxy)benzoate (2EC6F)
Yield: 0.173 g, 39.4 %. RF: 0.289 (100 % dichloromethane).
T$_{CrI}$ 146 °C T$_{N_FI}$ (132 °C)
$v_{max}$/cm$^{-1}$: 3061, 2969, 1727, 1705, 1601, 1584, 1526, 1487, 1474, 1463, 1411, 1343, 1275, 1251, 1230, 1194, 1175, 1159, 1143, 1112, 1094, 1076, 1047, 1017, 968, 949, 907, 891, 842, 824, 755, 736, 700, 686, 665, 631, 604, 572, 547, 520, 465, 450, 423, 409
$\delta_H$/ppm (400 MHz, CDCl$_3$): 8.14 (4 H, m, Ar-H), 7.37 (2 H, d, J 8.2 Hz, Ar-H), 7.29 (1 H, dd, J 11.4 Hz, 2.4 Hz, Ar-H), 7.22 (1 H, m, Ar-H), 6.96 (2 H, m, Ar-H), 3.97 (3 H, s, O-CH$_3$), 2.77 (2 H, quart, 7.6 Hz, Ar-CH$_2$-CH$_3$), 1.30 (3 H, t, 7.6 Hz, Ar-CH$_2$-CH$_3$)
$\delta_F$/ppm (376 MHz, CDCl$_3$): -113.38
$\delta_C$/ppm (100 MHz, CDCl$_3$): 164.63, 161.98, 161.94, 157.66, 156.88, 156.09, 155.99, 155.01, 149.92, 134.80, 134.73, 133.95, 130.49, 129.04, 127.27, 127.25, 126.36, 118.36, 118.32, 114.63, 114.03, 112.70, 112.47, 106.55, 56.49, 29.24, 15.35
MS = [M+Na]$^+$ : Calculated for C$_{23}$H$_{18}$FNO$_7$: 462.0965. Found: 462.0947. Difference: 3.9 ppm

### 6.3 3-Fluoro-4-nitrophenyl 2-methoxy-4-((4-propylbenzoyl)oxy)benzoate (3EC6F)
Yield: 0.196 g, 45.3 %. RF: 0.472 (100 % dichloromethane).
T$_{CrI}$ 150 °C T$_{N_FN_X}$ (116 °C) T$_{N_XN}$ (118 °C) T$_{NI}$ (123 °C)
$v_{max}$/cm$^{-1}$: 3061, 2962, 2928, 1727, 1705, 1602, 1583, 1527, 1488, 1473, 1410, 1344, 1275, 1251, 1231, 1194, 1178, 1158, 1113, 1094, 1078, 1052, 1018, 969, 948, 908, 891, 869, 841, 824, 753, 740, 700, 686, 670, 635, 609, 573, 546, 515, 460, 428, 407

δ$_H$/ppm (400 MHz, CDCl$_3$): 8.18 (1 H, t, J 8.8 Hz, Ar-H), 8.12 (3 H, m, Ar-H), 7.35 (2 H, d, J 8.4 Hz, Ar-H), 7.28 (1 H, dd, J 11.4 Hz, 2.4 Hz, Ar-H), 7.21 (1 H, m, Ar-H), 6.96 (2 H, m, Ar-H), 3.97 (3 H, s, O-CH$_3$), 2.70 (2 H, t, J 7.4 Hz, Ar-CH$_2$-CH$_2$-), 1.70 (2 H, m, Ar-CH$_2$-CH$_2$-CH$_3$), 0.97 (3 H, t, J 7.2 Hz, Ar-CH$_2$-CH$_2$-CH$_3$)

δ$_F$/ppm (376 MHz, CDCl$_3$): -113.34

δ$_C$/ppm (100 MHz, CDCl$_3$): 164.63, 161.98, 161.95, 157.66, 156.88, 156.09, 155.99, 155.01, 149.92, 134.80, 134.73, 133.95, 130.49, 129.04, 127.27, 127.25, 126.36, 118.36, 118.32, 114.63, 114.03, 112.70, 112.47, 106.58, 56.49, 38.27, 24.39, 13.89

MS = [M+H]$^+$ : Calculated for C$_{24}$H$_{21}$FNO$_7$: 454.1302. Found: 454.1311. Difference: 2.0 ppm

### 6.4 3-Fluoro-4-nitrophenyl 2-methoxy-4-((4-butylbenzoyl)oxy)benzoate (4EC6F)

Yield: 0.210 g, 56.3 %. RF: 0.400 (100 % dichloromethane).

T$_{CrI}$ 120 °C T$_{N_FN_X}$ (79 °C) T$_{N_XN}$ (92 °C) T$_{NI}$ (105 °C)

ν$_{max}$/cm$^{-1}$: 3062, 2959, 2931, 2857, 1727, 1706, 1601, 1585, 1526, 1487, 1474, 1462, 1413, 1343, 1275, 1253, 1230, 1195, 1178, 1159, 1143, 1114, 1093, 1075, 1054, 1017, 967, 950, 906, 891, 841, 826, 753, 700, 685, 671, 654, 635, 610, 592, 573, 547, 517, 464, 423, 408

δ$_H$/ppm (400 MHz, CDCl$_3$): 8.18 (1 H, t, J 8.9 Hz, Ar-H), 8.12 (3 H, m, Ar-H), 7.35 (2 H, d, J 8.4 Hz, Ar-H), 7.29 (1 H, dd, J 11.4 Hz, 2.4 Hz, Ar-H), 7.22 (1 H, m, Ar-H), 6.96 (2 H, m, Ar-H), 3.97 (3 H, s, O-CH$_3$), 2.72 (2 H, t, J 7.7 Hz, Ar-CH$_2$-CH$_2$-), 1.65 (2 H, m, Ar-CH$_2$-CH$_2$-CH$_2$-), 1.38 (2 H, m, Ar-CH$_2$-CH$_2$-CH$_2$-CH$_3$), 0.95 (3 H, t, J 7.3 Hz, Ar-CH$_2$-CH$_2$-CH$_2$-CH$_3$)

δ$_F$/ppm (376 MHz, CDCl$_3$): -113.34

δ$_C$/ppm (100 MHz, CDCl$_3$): 164.63, 161.98, 161.95, 157.66, 156.88, 156.09, 155.99, 155.01, 150.18, 134.80, 134.73, 133.95, 130.50, 128.98, 127.27, 127.25, 126.31, 118.36, 118.32, 114.63, 114.04, 112.70, 112.47, 106.58, 56.49, 35.96, 33.38, 22.46, 14.06

MS = [M+H]$^+$ : Calculated for C$_{25}$H$_{23}$FNO$_7$: 468.1459. Found: 468.1468. Difference: 1.9 ppm

### 6.5 3-Fluoro-4-nitrophenyl 2-methoxy-4-((4-pentylbenzoyl)oxy)benzoate (5EC6F)

Yield: 0.193 g, 45.8 %. RF: 0.474 (100 % dichloromethane).

T$_{CrI}$ 114 °C T$_{N_XN}$ (85 °C) T$_{NI}$ (109 °C)

ν$_{max}$/cm$^{-1}$: 3062, 2958, 2927, 2854, 1729, 1706, 1602, 1584, 1527, 1488, 1474, 1412, 1343, 1275, 1253, 1231, 1194, 1177, 1159, 1114, 1094, 1073, 1054, 1023, 968, 949, 906, 891, 864, 841, 825, 750, 728, 700, 686, 671, 646, 635, 610, 573, 547, 516, 481, 465, 429, 420, 409

δ$_H$/ppm (400 MHz, CDCl$_3$): 8.18 (1 H, t, J 9.0 Hz, Ar-H), 8.12 (3 H, m, Ar-H), 7.34 (2 H, d, J 8.4 Hz, Ar-H), 7.29 (1 H, dd, J 11.4 Hz, 2.4 Hz, Ar-H), 7.21 (1 H, m, Ar-H), 6.96 (2 H, m, Ar-H), 3.97 (3 H, s, O-CH$_3$), 2.71 (2 H, t, J 7.5 Hz, Ar-CH$_2$-CH$_2$-), 1.67 (2 H, m, Ar-CH$_2$-CH$_2$-CH$_2$-), 1.35 (4 H, m, Ar-CH$_2$-CH$_2$-CH$_2$-CH$_2$-CH$_3$), 0.91 (3 H, t, J 6.9 Hz, Ar-CH$_2$-CH$_2$-CH$_2$-CH$_2$-CH$_3$)

δ$_F$/ppm (376 MHz, CDCl$_3$): -113.33

δ$_C$/ppm (100 MHz, CDCl$_3$): 164.63, 161.98, 161.95, 157.67, 156.89, 156.10, 155.99, 155.02, 150.21, 134.81, 134.74, 133.95, 130.51, 128.99, 127.27, 127.25, 126.31, 118.36, 118.33, 114.63, 114.04, 112.71, 112.47, 106.58, 56.49, 36.24, 31.56, 30.95, 22.65, 14.16

MS = [M+H]$^+$ : Calculated for C$_{26}$H$_{25}$FNO$_7$: 482.1615. Found: 482.1636. Difference: 4.4 ppm

### 6.6 3-Fluoro-4-nitrophenyl 2-methoxy-4-((4-hexylbenzoyl)oxy)benzoate (6EC6F)

Yield: 0.169 g, 40.5 %. RF: 0.314 (100 % dichloromethane).

T$_{CrI}$ 104 °C T$_{N_XN}$ (69 °C) T$_{NI}$ (97 °C)

$v_{max}$/cm$^{-1}$: 3036, 2927, 2853, 1728, 1708, 1602, 1585, 1527, 1487, 1474, 1412, 1344, 1309, 1276, 1254, 1229, 1195, 1176, 1157, 1143, 1113, 1094, 1073, 1054, 1024, 1016, 967, 949, 906, 890, 851, 843, 823, 752, 702, 686, 671, 644, 636, 611, 594, 572, 547, 517, 464, 439, 421, 408

$\delta_H$/ppm (400 MHz, CDCl$_3$): 8.18 (1 H, t, J 8.7 Hz, Ar-H), 8.12 (3 H, m, Ar-H), 7.34 (2 H, d, J 8.3 Hz, Ar-H), 7.29 (1 H, dd, J 11.4 Hz, 2.4 Hz, Ar-H), 7.21 (1 H, m, Ar-H), 6.96 (2 H, m, Ar-H), 3.79 (3 H, s, O-CH$_3$), 2.72 (2 H, t, J 7.7 Hz, Ar-CH$_2$-CH$_2$-), 1.65 (2 H, m, Ar-CH$_2$-CH$_2$-CH$_2$-), 1.33 (6 H, m, Ar-CH$_2$-CH$_2$-CH$_2$-CH$_2$-CH$_2$-CH$_3$), 0.90 (3 H, t, J 6.8 Hz, Ar-CH$_2$-CH$_2$-CH$_2$-CH$_2$-CH$_2$-CH$_3$)

$\delta_F$/ppm (376 MHz, CDCl$_3$): -113.37

$\delta_C$/ppm (100 MHz, CDCl$_3$): 164.62, 161.99, 161.95, 157.67, 156.90, 156.11, 156.01, 155.02, 150.21, 134.83, 134.76, 133.94, 130.51, 128.98, 127.26, 127.24, 126.34, 118.35, 118.31, 114.68, 114.04, 112.69, 112.46, 106.60, 56.49, 36.27, 31.79, 31.21, 29.05, 22.72, 14.22

MS = [M+H]$^+$ : Calculated for C$_{27}$H$_{27}$FNO$_7$: 496.1772. Found: 496.1771. Difference: 0.2 ppm

## Additional Experimental Data

**Table S7**. The phase transition temperatures for the *n*EC6F series with an alkyl (C*n*) terminal chain, the temperatures are given in °C and associated transition enthalpy changes (in parenthesis) in Jg$^{-1}$

| n | m.p. | Phase Sequence |
| --- | --- | --- |
| 1 | 169.0(97.0) | N$_F$-155.8(11.6)-Iso |
| 2 | 145.9(90.6) | N$_F$-132.1(9.3)-Iso |
| 3 | 150.0(106.2) | N$_F$-116-N$_X$-118.0(0.8)-N-122.8(2.4)-Iso |
| 4 | 120.2(78.6) | N$_F$-79.0(0.6)-N$_X$-92.2(0.03)-N-105.4(2.0)-Iso |
| 5 | 113.8(76.4) | N$_X$-84.9(0.02)-N-109.4(2.2)-Iso |
| 6 | 104.2(59.4) | N$_X$-68.5(0.02)-N-97.2(1.4)-Iso |

**Table S8**. The phase transition temperatures for the *n*OEC3F series with an alkyloxy (OC*n*) terminal chain, the temperatures are given in °C and associated transition enthalpy changes (in parenthesis) in Jg$^{-1}$

| n | m.p. | Phase Sequence |
| --- | --- | --- |
| 1 | 179.0(113.4) | N$_F$-142.4(2.4)-N-156.8(2.0)-Iso |
| 2 | 164.0(103.4) | N$_F$-104.2(0.6)-N$_X$-117.9(0.03)-N-152.5(2.4)-Iso |
| 3 | 144.9(90.2) | N$_X$-79.0(0.01)-N-133.3(1.4)-Iso |
| 4 | 127.8(87.7) | N$_X$-38-N-130.3(1.5)-Iso |
| 5 | 91.9(62.5) | N-118.0(1.1)-Iso |

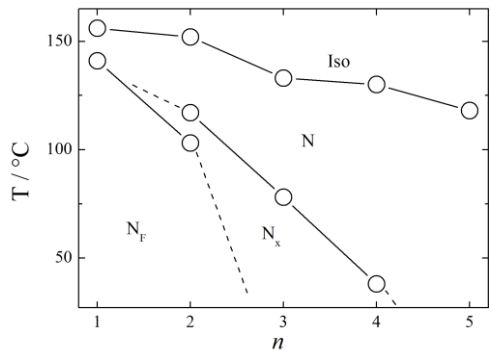

**Figure S1**. Phase diagram for studied homologue series of ferronematogens with alkyloxy terminal chain, *n*, molecular formula given above.

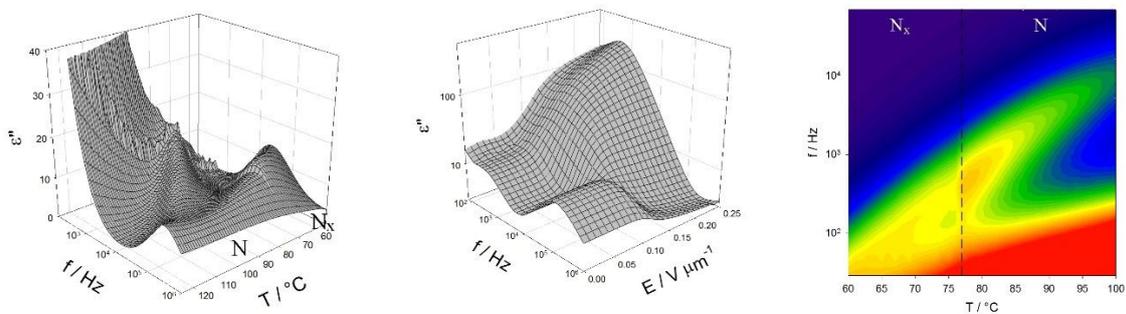

**Figure S2**. The Imaginary part of the dielectric susceptibility measured for homologue 3OEC3F: (left) the temperature and frequency dependence across the N-$N_X$ phase sequence, (middle) the frequency and bias field dependence in the $N_X$ phase (at T = 73 °C); where the high-frequency mode is suppressed and the lower-frequency 'ferroelectric' mode is excited above threshold field of 0.11 V/µm; and (right) a map showing evolution the 'ferroelectric' mode vs. temperature and frequency, the measurements were performed under bias electric field 0.3 V/µm.

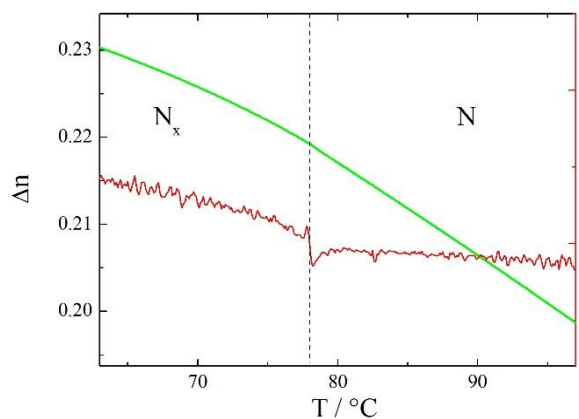

**Figure S3**. Optical birefringence (green line) of 3OEC3F measured with green light (λ = 532 nm) across the N-$N_X$ phase transition. The red line shows derivative d(Δn)\dT, evidencing the phase transition temperature.

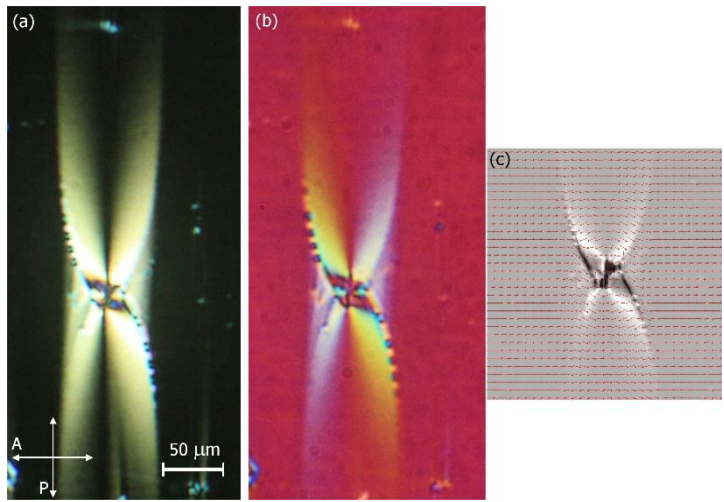

**Figure S4**. Optical texture with focal conic-twin like defects in $N_F$ phase taken for 4EC6F in 1.6-µm-thick cell with planar anchoring: (a) between crossed polarizers, (b) with λ retardation plate inserted at 45 deg. with respect to polarizers, (c) the optical retardation and director field determined with Abrio system (at λ = 546 nm). The defects are anchored on rods used as separators in glass cells.